\documentclass[12pt]{article}
\usepackage{palatino}
\usepackage{a4}
\usepackage{axodraw}
\usepackage{cite}
\usepackage{graphicx}
\usepackage{amssymb}
\usepackage{amsmath}
\usepackage{epsfig}
\usepackage[tmargin=1cm,bmargin=1cm,lmargin=2cm,rmargin=2cm]{geometry}

\newcommand{\qb}{\ensuremath{\overline q}}

\newcommand{\ixotomds}{\ensuremath{ \int_{x_1}^{1-\delta_s} \frac{dz}{z} {\cal H}(z,\e,\delta_c)  }}

\newcommand{\ixttomds}{\ensuremath{ \int_{x_2}^{1-\delta_s} \frac{dz}{z} {\cal H}(z,\e,\delta_c)  }}
\newcommand{\ixtto}{\ensuremath{ \int_{x_2}^{1} \frac{dz}{z} {\cal H}(z,\e,\delta_c)  }}

\newcommand{\be}{\begin{equation}}
\newcommand{\bdm}{\begin{displaymath}}
\newcommand{\bea}{\begin{eqnarray}}
\newcommand{\beastar}{\begin{eqnarray*}}

\newcommand{\ee}{\end{equation}}

\newcommand{\e}{\ensuremath{\epsilon}}

\newcommand{\edm}{\end{displaymath}}
\newcommand{\eea}{\end{eqnarray}}
\newcommand{\eeastar}{\end{eqnarray*}}

\newcommand{\Msq}{\ensuremath{ { \overline {|M^{(0)}|^2} }}}

\newcommand{\rarrow}{\ensuremath{\rightarrow}}

\nopagebreak

\begin{document}
\baselineskip 22pt

\begin{flushright}
\end{flushright}
\vskip 65pt
\begin{center}
{\large \bf
Direct photon pair production at the LHC to ${\cal O}(\alpha_s)$ \\in TeV scale gravity models
} 
\\
\vspace{8mm}
{\bf
M.C.~Kumar$^{a,b}$
\footnote{mc.kumar@saha.ac.in},
Prakash Mathews$^b$
\footnote{prakash.mathews@saha.ac.in},
}\\
{\bf
V. Ravindran$^c$
\footnote{ravindra@hri.res.in},
Anurag Tripathi$^c$
\footnote{anurag@hri.res.in}
}\\
\end{center}
\vspace{10pt}
\begin{flushleft}
{\it
a)~~Saha Institute of Nuclear Physics, 1/AF Bidhan Nagar, Kolkata 700 064, India.\\
b)~~School of Physics, University of Hyderabad, Hyderabad 500 046, India.\\
c)~~Regional Centre for Accelerator-based Particle Physics,\\
~~~~Harish-Chandra Research Institute,
 Chhatnag Road, Jhunsi, Allahabad, India.\\
}
\end{flushleft}
\vspace{10pt}
\centerline{\bf Abstract}
The first results on next-to-leading order QCD corrections to production of direct photon pairs
in hadronic collisions in the extra dimension models--- ADD and RS are presented.
Various kinematical distributions are obtained to order $\alpha_s$ in QCD by 
taking into account all the parton level subprocesses.  Our Monte Carlo based code
incorporates all the experimental cuts suitable for physics studies at the LHC.
We estimate the impact of the QCD corrections on various observables and find that they
are significant.  We also show the reduction in factorisation scale uncertainity when
${\cal O}(\alpha_s)$ effects are included.

\vskip12pt
\vfill
\clearpage

\setcounter{page}{1}
\pagestyle{plain}

\section { Introduction}
The hierarchy problem is a long standing problem and has been the 
main motivation for physics beyond the SM.  The hierarchy between 
the electroweak scale and the Planck scale has in the past been 
addressed 
by modifying the particle content of the 
theory--- supersymmetry and technicolor belong to this category.  A 
paradigm shift in this approach was proposed by Arkani-Hamed, 
Dimopoulos and Dvali (ADD) \cite{add}, wherein they modified only the 
gravity sector. 
Though the idea of extra dimensions existed since the 1900s, all models 
assumed that gravity together with 
other 
interactions could live in the full extra dimensions.  Consistency with 
the experimental observations, demands that these extra dimensions be very small.  The 
ADD scenario explored the possibility of allowing only gravity to 
probe all dimensions and studied the constraints on the size of 
the extra dimensions.  It turns out that for more then one extra 
dimensions, their size could be large without contradicting any known 
experimental observations and consequently explain the weakness of 
gravity in 4-dimensions.  An alternate solution of the hierarchy problem 
was suggested by Randall and Sundrum \cite{rs} with a single extra 
dimension in an Anti-de-Sitter (AdS$_5$) metric.  The ADD and RS models 
used the geometry of the extra dimensions to account for the 
hierarchy between the electroweak scale and the Planck scale.
In order that these models are consistent with present experimental 
length scale probed at colliders, 
it is essential that these extra dimensions remain hidden.  Various 
extra dimensional models have used different physical mechanisms to hide 
them.  The extra dimensions could be small and compact 
wherein all the SM fields are allowed to propagate, or alternatively the 
brane world scenarios where the SM particles are confined to the 
brane.
In the brane world scenarios the extra dimensions could be of macroscopic size (ADD)
without contradiction with present experiments.  ADD and RS are both 
brane-world scenarios.


In the ADD case the compactified extra dimensions are flat
and could be large.  
It follows from Gauss Law that 
the effective Planck scale $M_P$ in 4-dimension is related to a
fundamental scale $M_S$ in $4+d$-dimension through the volume of the
compactified $d$ extra spatial dimensions \cite{add}.  
The hierarchy between the Planck scale and the electroweak scale ($M_W$)
is solved by assuming $M_S \approx M_W$.
A viable mechanism to hide the 
extra spatial dimensions, is to introduce a 3-brane with negligible 
tension and localise the SM particles on it.  
Only gravity is allowed to propagate in all dimensions.  
The $d$ spatial dimensions are compactified on a torus of common circumference $R$.
The spectrum consists of the SM fields and 
a tower of Kaluza-Klein (KK) modes of the graviton fields.
The number of extra spacial dimension possible is $d \ge 2$ from current 
experimental limits on deviation from inverse square law \cite{expt}.  

This was the first extra dimension model in which 
the compactified dimensions could be of macroscopic size and consistent 
with present experiments.  In this model, new physics can appear at a 
mass scale of the order of a TeV.  
The interaction of the KK modes $h_{\mu\nu}^{(\vec n)}$ with the SM fields
localised on the 3-brane is given by
\begin{eqnarray}
{\cal L} = - \frac{\kappa}{2} \sum_{\vec n=0}^\infty T^{\mu\nu} (x)
                      ~h_{\mu\nu}^{(\vec n)} (x) ~,
\end{eqnarray}
where $\kappa=\sqrt{16 \pi}/M_P$ and $T^{\mu\nu}$ is the energy-momentum 
tensor of the SM fields on the 3-brane.  The zero mode corresponds to the 
usual 4-dimensional massless graviton.  The KK modes are $M_P$ 
suppressed but the high multiplicity could lead to observable effects.  
In a process involving a virtual exchange of KK modes from SM particles, 
the sum of KK propagators ${\cal D} (Q^2)$ is given by 
\begin{eqnarray}
\kappa^2 {\cal D} (Q^2) &=& \kappa^2 \sum_n \frac{1}{Q^2-m_n^2 +i \epsilon}
~,
\nonumber\\
&=& \frac{8 \pi}{M_S^4} \left(\frac{Q}{M_S}\right)^{(d-2)} 
\Big[ -i \pi + 2 I(\Lambda/Q) \Big] ~,
\end{eqnarray}
the integral $I(\Lambda/Q)$ is a result of the summation over the
non-resonant KK modes and the term proportional to $\pi$ is due to the 
resonant production of a single KK modes \cite{hlz}.  $\Lambda$ is 
the explicit cut-off on the KK sum which is identified with the scale 
of the extra dimension theory $M_S$ \cite{grw,hlz}.  The $\kappa^2$
suppression in a virtual exchange is compensated for by the high 
multiplicity, after the KK modes are summed over.  
The ADD scenario raises 
the exciting possibility of observing quantum gravity at the LHC.  
Basic collider signals of the ADD scenario could be (a) real KK mode 
production resulting in missing energy in association with a SM 
gauge boson or a hadronic jet or (b) virtual KK mode exchange which
could lead to deviations from the SM predictions.  Interesting 
phenomenological consequence have been considered in \cite{grw,hlz,
peskin,mrs}.


In the RS model there is only one extra spacial dimension and the extra
dimension is compactified to a circle of circumference $2 L$ and further
orbifolded by identifying points related by $y \to -y$.  Two brane are
placed at orbifold fixed points, $y=0$ with positive tension called the
Planck brane and a second brane at $y=L$ with negative tension called the
TeV brane.  For a special choice of parameters, it turns out that the
5-dimensional Einstein equations have a warped solution for $0<y<L$ with
metric $g_{\mu\nu} (x^\rho,y)=\exp(-2 k y) ~\eta_{\mu\nu}$, $g_{\mu y}=0$
and $g_{y y}=1$.  This space is not factorisable and has a constant negative
curvature--- $AdS_5$ space-time.  $k$ is the curvature of the $AdS_5$
space-time and $\eta_{\mu\nu}$ is the usual 4-dimensional flat Minkowski
metric.  In this model the mass scales vary with $y$ according to the
exponential warp factor.  If gravity originates on the brane at $y=0$,
TeV scales can be generated on the brane at $y=L$ for $k L \sim 10$.
The apparent hierarchy is generated by the exponential warp
factor and no additional large hierarchies appear.  The size of the
extra dimension is of the order of $M_P^{-1}$.  Further it has been
showed that \cite{gw} the value of $k L$ can be stabilised without
fine tuning by minimising the potential for the modulus field
which describes the relative motion of the two branes.  In the RS
model graviton and the modulus field can propagate the full
5-dimensional space time while the SM is confined to the TeV brane.
The 4-dimensional spectrum contains the KK modes, the zero mode is
$M_P$ suppressed while the excited modes are massive and are only TeV
suppressed.  The mass gap of the KK modes is determined by the difference
of the successive zeros of the Bessel function $J_1 (x)$ and the scale
$m_0=k ~e^{-\pi k L}$.
As in the ADD case the phenomenology of the RS model concerns
the effect of massive KK modes of the graviton, though the spectrum
of the KK mode is quite different.

In the RS model the massive KK modes $h_{\mu\nu}^{(n)} (x)$ interacts
with the SM fields
\begin{eqnarray}
{\cal L}_{int} \sim - \frac{1}{M_P} T^{\mu\nu} ~ h_{\mu\nu}^{(0)} 
                 - \frac{1}{M_P ~e^{-\pi k L}} \sum_{n=1}^\infty T^{\mu\nu} 
                   ~ h_{\mu\nu}^{(n)} ~,
\end{eqnarray}
where the energy-momentum tensor $T^{\mu\nu}$ of the SM fields lives on the
3-brane at $y=L$ and its coupling to the massive KK modes could be TeV
suppressed if $k L \sim 10$.  The masses of $h_{\mu\nu}^{(n)}$ are given 
by $M_n=x_n ~k ~e^{-\pi k L}$, where $x_n$ are the zeros of the Bessel 
function $J_1 (x)$.  In this model there are two parameters which are 
$c_0=k/M_P$, the effective coupling and $m_1$ the mass of the first KK 
mode.  Except for an overall warp factor the Feynman rules of RS are the 
same as those of the ADD model.  In the RS case since the spectrum is 
massive and its spacing is determined by $x_n$, the summation of the 
propagator is given by \cite{dhr}
\begin{eqnarray}
{\cal D} (Q^2) &=& \sum_n \frac{1}{s -M_n^2 +i M_n \Gamma_n}
~,
\nonumber\\
&=& \frac{1}{m_0^2} \sum_n \frac{X^2-X_n^2 - i \frac{\Gamma_n}{m_0} X_n}
{(X^2-X_n^2)^2 + \frac{\Gamma^2_n}{m^2_0} X^2_n} ~,
\end{eqnarray}
where $X=\sqrt{s}/m_0$ and $X_n=M_n/m_0$.  The summation over $n$ is
kinematically bounded.  Further the RS KK mode of mass $M_n$ if 
decays only to SM particles, the decay width $\Gamma_n$ is fixed. 
The signal is now resonant enhancement over the SM predictions and is 
very distinct compared to the ADD case which leads to an enhancement 
of the tail for say the invariant mass distribution.   

Production of photon pairs at hadronic colliders is an important
process as it provides a clean channel not only to test the predictions
of the SM but also of any new physics beyond it.
An extensive study of this process exists in the literature \cite{TwoPhotonBkgd1} 
in the context of light Higgs-boson searches as a light 
Higgs Boson decays dominantly to two photons. This channel has 
also been widely used for various beyond SM studies \cite{BSMdiphoton}. Recently 
we completed a next-to-leading order computation for this process
in the context of theories with large extra-dimension and unparticle model \cite{ourpapers1}.
The present paper aims at a full next-to-leading order computation for
production of isolated direct photon pairs at the LHC at $\sqrt{S} =14$~TeV,
and to obtain various
kinematical distributions with experimental cuts imposed on the 
photons. 
All the details of the calculation are presented here which
were not presented in previous publications \cite{ourpapers1,ourpapers2}.
The estimate of enhancement over LO
result and the improvement in scale uncertainties in going from a
LO result to a NLO result are 
the main motivations for this work.  In \cite{drellyan}, QCD radiative
corrections beyond LO to Drell-Yan process in gravity mediated models
were first studied and it was found to be large.
Subsequently, they were used (see \cite{Abulencia:2006kk,Abazov:2007ra})
to constrain the model parameters.

The NLO calculation
presented here uses both analytical and Monte Carlo integration 
methods. It is easy to impose experimental cuts in a Monte Carlo 
calculation than a fully analytical computation. Our code is 
based on the method of { \it two cutoff phase space slicing } \cite{twocutoff} to
deal with various singularities appearing in the NLO computation
and to implement the numerical integrations over phase space.
It has been applied to diphoton production in \cite{Bailey:1992br}.  
This method is nicely reviewed in \cite{twocutoff-rev}.
All the analytical results presented in this paper were evaluated using the algebraic 
manipulation program FORM \cite{FORM}.

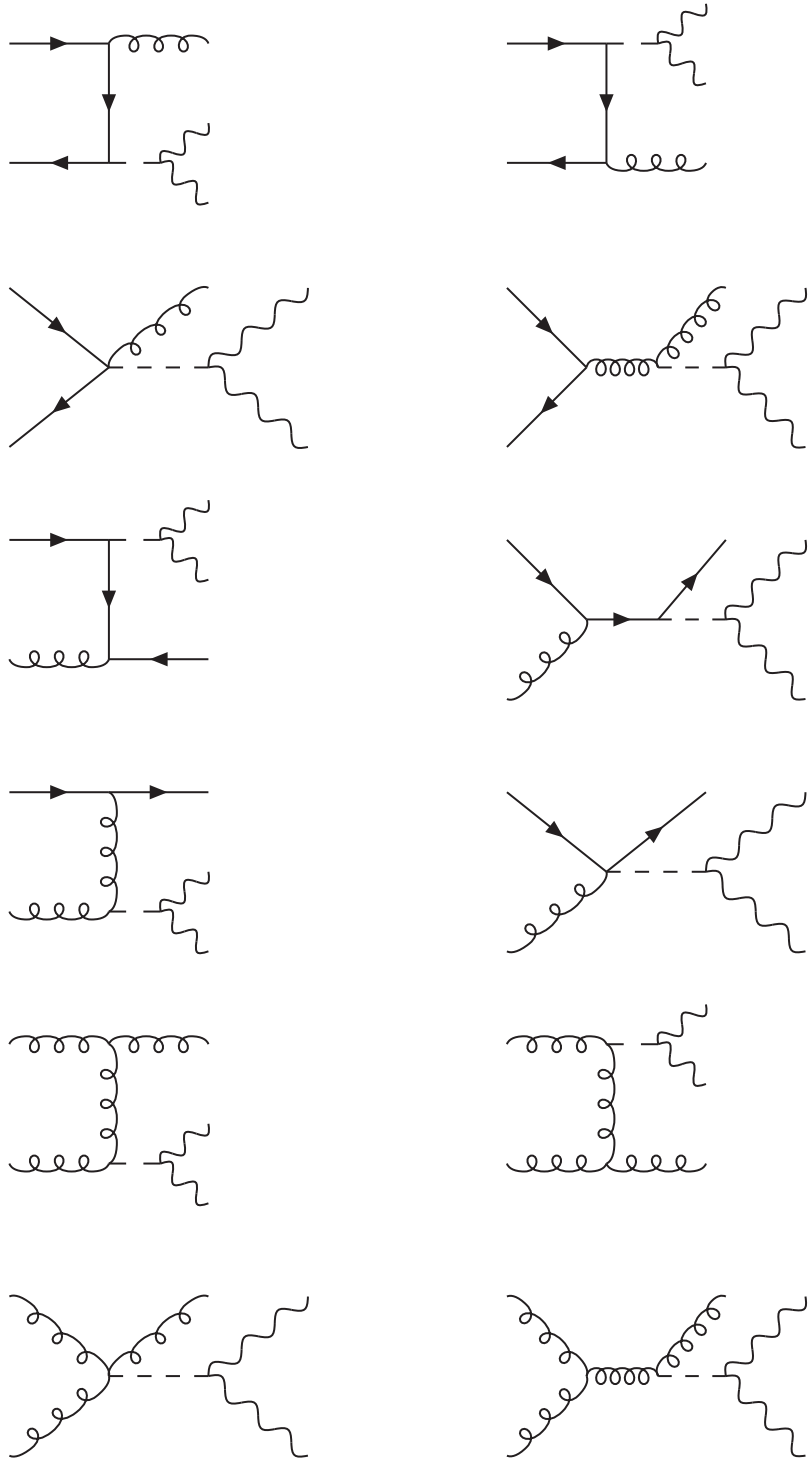
\begin{figure}
\SetScale{1.5}
\noindent
\begin{center}
\hspace{-5cm}
\begin{picture}(200,40)
\ArrowLine(0,40)(25,40)
\ArrowLine(25,40)(25,10)
\ArrowLine(25,10)(0,10)
\Gluon(25,40)(50,40){2}{3}
\DashLine(25,10)(38,10){4}
\Photon(38,10)(50,20){2}{2}
\Photon(38,10)(50,0){2}{2}
\ArrowLine(125,40)(150,40)
\ArrowLine(150,40)(150,10)
\ArrowLine(150,10)(125,10)
\Gluon(175,10)(150,10){2}{3}
\DashLine(150,40)(163,40){4}
\Photon(163,40)(175,50){2}{2}
\Photon(163,40)(175,30){2}{2}

\end{picture}
\end{center}
\vspace{1.0cm}
\begin{center}
\hspace{-5cm}
\begin{picture}(200,40)
\ArrowLine(0,40)(25,20)
\ArrowLine(25,20)(0,0)
\Gluon(25,20)(50,40){2}{3}
\DashLine(25,20)(50,20){4}
\Photon(50,20)(75,40){2}{3}
\Photon(50,20)(75,0){2}{3}
\ArrowLine(125,40)(145,20)
\ArrowLine(145,20)(125,0)
\Gluon(145,20)(163,20){2}{4}
\Gluon(163,20)(180,40){2}{4}
\DashLine(163,20)(180,20){4}
\Photon(180,20)(200,40){2}{3}
\Photon(180,20)(200,0){2}{3}
\end{picture}
\end{center}
\vspace{1.0cm}
\begin{center}
\hspace{-5cm}
\begin{picture}(200,40)
\ArrowLine(0,40)(25,40)
\ArrowLine(25,40)(25,10)
\Gluon(25,10)(0,10){2}{3}
\ArrowLine(50,10)(25,10)
\DashLine(25,40)(38,40){4}
\Photon(38,40)(50,50){2}{2}
\Photon(38,40)(50,30){2}{2}
\ArrowLine(125,40)(145,20)
\Gluon(145,20)(125,0){2}{3}
\ArrowLine(145,20)(163,20)
\ArrowLine(163,20)(180,40)
\DashLine(163,20)(180,20){4}
\Photon(180,20)(200,40){2}{3}
\Photon(180,20)(200,0){2}{3}
\end{picture}
\end{center}
\vspace{1.0cm}
\begin{center}
\hspace{-5cm}
\begin{picture}(200,40)
\ArrowLine(0,40)(25,40)
\Gluon(25,40)(25,10){2}{3}
\Gluon(25,10)(0,10){2}{3}
\ArrowLine(25,40)(50,40)
\DashLine(25,10)(38,10){4}
\Photon(38,10)(50,20){2}{2}
\Photon(38,10)(50,0){2}{2}
\ArrowLine(125,40)(150,20)
\Gluon(150,20)(125,0){2}{3}
\ArrowLine(150,20)(175,40)
\DashLine(150,20)(175,20){4}
\Photon(175,20)(200,40){2}{3}
\Photon(175,20)(200,0){2}{3}
\end{picture}
\end{center}
\vspace{1.0cm}
\begin{center}
\hspace{-5cm}
\begin{picture}(200,40)
\Gluon(0,40)(25,40){2}{3}
\Gluon(25,40)(25,10){2}{3}
\Gluon(25,10)(0,10){2}{3}
\Gluon(25,40)(50,40){2}{3}
\DashLine(25,10)(38,10){4}
\Photon(38,10)(50,20){2}{2}
\Photon(38,10)(50,0){2}{2}
\Gluon(125,40)(150,40){2}{3}
\Gluon(150,40)(150,10){2}{3}
\Gluon(150,10)(125,10){2}{3}
\Gluon(175,10)(150,10){2}{3}
\DashLine(150,40)(163,40){4}
\Photon(163,40)(175,50){2}{2}
\Photon(163,40)(175,30){2}{2}
\end{picture}
\end{center}
\vspace{1.0cm}
\begin{center}
\hspace{-5cm}
\begin{picture}(200,40)
\Gluon(0,40)(25,20){2}{3}
\Gluon(25,20)(0,0){2}{3}
\Gluon(25,20)(50,40){2}{3}
\DashLine(25,20)(50,20){4}
\Photon(50,20)(75,40){2}{3}
\Photon(50,20)(75,0){2}{3}
\Gluon(125,40)(145,20){2}{3}
\Gluon(145,20)(125,0){2}{3}
\Gluon(145,20)(163,20){2}{4}
\Gluon(163,20)(180,40){2}{4}
\DashLine(163,20)(180,20){4}
\Photon(180,20)(200,40){2}{3}
\Photon(180,20)(200,0){2}{3}
\end{picture}
\end{center}
\caption{Real emission contributions}
\label{real}
\end{figure}

At the lowest order in $ \alpha_s $ ie., at $ \alpha_s ^0 $, two
photons in the final state are produced in quark anti-quark 
annihilation subprocess $ q \qb \rarrow \gamma \gamma $ in the
SM. For low invariant mass photon pairs $ gg \rarrow \gamma \gamma $,
although of order $ \alpha_s^2 $, is comparable to  
$ q \qb \rarrow \gamma \gamma $. This is due to the large 
gluon densities at small $x$. In light Higgs boson search studies
this subprocess plays an important role, and it is treated formally
as a leading order contribution although it is of order $ \alpha_s^2$
\cite{TwoPhotonBkgd1,gg} and is really a next-to-next-to leading order contribution.
However, it falls rapidly with
increasing invariant mass and in the mass range of interest for the TeV scale gravity models,
it need not be included at LO. 
We have demonstrated in \cite{ourpapers2} that this
subprocess in the SM is few orders of magnitude smaller than 
that of $q\bar{q}$ when $Q > 500~$GeV.
This subprocess amplitude can interfere with the gluon initiated LO subprocess
in ADD (also in RS) giving order $\alpha_s$ contribution which is included
in our study.  In addition   
order $\kappa^2$ gravity mediated Feynman diagrams fig.(\ref{born})
$ q q \rarrow \gamma \gamma $ and $ g g \rarrow \gamma \gamma $ 
also appear at the leading order. The NLO computation involves
two kinds of matrix elements. 
\begin{itemize}
\item Virtual diagrams with loops which contribute through their 
      interference with the LO diagrams (see fig.(\ref{virt})). 
\item Real emission diagrams with an additional parton in the final
      state (see fig.(\ref{real})).
\end{itemize} 
Both the virtual and real corrections have been evaluated with
5 quark flavors and in the limit of vanishing of quark masses.
The $n$-point tensor integrals appearing from integration 
over loop-momenta were simplified using Passarino-Veltman reduction and
computation was carried with dimensional regularization using
$n=4+\epsilon$, and divergences were subtracted or factorized 
in ${\overline {MS}}$ scheme.

Photons not only arise directly in a parton subprocess but
also through fragmentation of a parton into a photon and a
jet of hadrons collinear to it. This fragmentation is a 
non-perturbative phenomenon. A parton level computation involving
only direct photons without including fragmentation photons is
plagued with QED collinear singularities.  It is evident from
the SM Feynman diagrams.  This singularity can be
absorbed into fragmentation functions describing probability of
a parton fragmenting into a photon in much the same way as 
initial state collinear singularities are absorbed in parton
distribution functions.
However, fragmentation functions are not known to a good 
accuracy.  An alternative is to avoid these fragmentation functions
and simultaneously suppress final state QED singularity by using the
{\it smooth cone isolation } criterion advocated by Frixione \cite{Frixione:1998jh}.

{ \it Smooth cone isolation}: The aim of this isolation criterion
is to suppress the final state QED collinear singularities and at the 
same time also remove the fragmentation photons in an infrared 
safe manner. Let the $z$-axis coincide with the proton-proton 
collision line and $\theta$ and $\phi$ denote the polar and 
azimuthal angles respectively. It is, however, more convenient 
to use the pseudo-rapidity in the context of hadron colliders, as
they are additive under boosts.  
The fragmentation photons are embedded in hadronic jets and the 
prescription to isolate a photon from hadronic activity
is to draw concentric circles around it in $\eta-\phi$ plane 
with the largest circle having a fixed radius $R_0$,
which will be taken as $R_0=0.4$,
and demand that the sum of hadronic transverse energy in any circle
of radius $R<R_0$ be less than some specified amount $ H(R)$.
Thus as we move closer to the photon lesser hadronic energy is 
allowed in its neighborhood.
In order
that this criterion does not disturb cancellation of infrared singularities
$H(R)$ is restricted to have the limit $H(R \rarrow 0) =0$.
Here we would use the following choice for $H(R)$
\be
H(R) = E_T^{iso} \left( \frac{1-\cos R}{1-\cos R_0} \right)^n\,,
\label{HR}
\ee
where $E_T^{iso}$ is a fixed energy. 
In this paper, for our analysis, $n=2$ and $E_T^{iso}=15$ GeV would be default
choices. 

The paper is organized as follows. In section-2, we outline the next-to-leading
order computation in the two cut off phase space slicing method and present all
the analytical results that go into our Monte Carlo code with the exception of
$2 \rightarrow 3$ subprocess matrix elements
\footnote{These can be obtained from us on request}.    
In section-3, the numerical results and discussion on various kinematical distributions
are presented.  Finally we conclude in the section-4.  

\section {Outline of computation}
\subsection{ Leading order processes}
A parton level $2 \rarrow 2$ process at the leading order is
of the generic form
\be
a(p_1) + b(p_2) \rarrow  \gamma(p_3)+ \gamma(p_4).
\ee
where $a$ and $b$ are either quark and anti-quark or gluons.
The exact matrix elements in $n=4+\e$ dimensions  for $q \qb $ and $gg$ 
initiated subprocesses are
\bea
\Msq_{q \qb, sm} &=& \frac{e_q^4}{N} 
       ~\left[ \frac{u}{t} + \frac{t}{u} 
              + \e \left(1 +\frac{u}{t}  +\frac{t}{u} \right)
              + \frac{\e^2 }{4} \left( 2 + \frac{u}{t} + \frac{t}{u} \right)
       \right] 
\\[2ex]
\Msq_{q \qb, int} &=& - ~{ \kappa^2 {\cal R}e {\cal D}(s)~}~ {e_q^2 \over 8 N}
                    ~\Big[4 \left( t^2 +u^2 \right)
                          + \e \left( 3t^2 +3u^2 +2ut \right)
                    \Big] 
\\[2ex]
\Msq_{q\qb, gr}  &=& \frac{\kappa^4 |{\cal D}(s)|^2}{ 16 N}
                  ~\Big[
                       ut^3 +tu^3 + \frac{\e}{4} \left( 3tu^3 +3t^3u +2u^2t^2 \right)
                  \Big],
\\[2ex]
\Msq_{gg,gr} &=&  \frac{\kappa^4 |{\cal D}(s)|^2}{N^2-1} 
~\Big [ \frac{81}{128 (3+\e)^2}  s^4  
 + \frac{27}{64 (3+\e)} s^2 \left( u^2 +14t u +t^2 \right)  
\nonumber\\[0.1ex]
&& + \frac{5}{2 (2+\e)^2} s^2 t u
   - \frac{1}{16 (2+\e)}  s^2 \left( 7 u^2+94 t u + 7 t^2 \right)
\nonumber\\[0.1ex]
&&   + \frac{1}{128} \left( 9 t^4 +28 t^3 u + 54 t^2 u^2 +28 t u^3 +9 u^4 \right) 
\Big]
\eea
where $sm,~gr,~int$ represent contributions from SM, gravity, and interference of
SM with gravity induced process respectively,  
$s,t,u$ are the usual Mandelstam invariants, $e_q$ is the charge of a 
quark or anti-quark and $\kappa$ is the coupling of gravity to SM fields.
The bar over the symbol $M$ 
represents that the matrix elements have been averaged over initial
helicities and color, and summed over the final ones.  
A factor of 1/2 has been included for identical final state
photons. This expression has been evaluated for quarks with $N$ and
gluons with $N^2-1$ color degrees of freedom.
\subsection{ Virtual process}
The order $\alpha_s$ corrections to leading order process 
come from interference between Born graphs and 
virtual graphs. It is to be noted that 
the virtual contribution here does not contain UV singularities. The
reason lies in the facts that (i) electromagnetic coupling $\alpha$
does not receive any QCD corrections, (ii) and that the gravitons
couple to the energy momentum tensor of SM fields which is a 
conserved quantity and does not get renormalized. 
The Feynman diagrams with external leg corrections are not shown
as these vanish in the dimensional regularization in the massless
limit. We give below the order $\alpha_s$ squared matrix element 
coming from virtual processes.  The SM contribution is found to be
\bea
\overline{|M^{V}|^2}_{q \qb,sm} 
  &=&
       a_s(\mu_R^2) f(\e,\mu_R^2,s) C_F ~\Bigg[~ 
       \Upsilon \left(\epsilon \right)
      ~  
            {\Msq}_{q \qb, sm}
          +      2 \frac{e_q^4}{N}~ 
                  \Big\{  (4\zeta(2) -7)\frac{u}{t} 
\nonumber\\[1ex]
&&+  \left(2 + 3\frac{u}{t} \right) \ln \frac{-t}{s}   
                         +  \left(2 +\frac{u}{t} + 2\frac{t}{u} \right) \ln^2 \frac{-t}{s} 
                    + t \leftrightarrow u\Big\}\Bigg],  
\eea
the interference of SM with the gravity mediated processes are 
\bea
\overline{|M^{V}|^2}_{q \qb,int} 
  &=&
       a_s(\mu_R^2) f(\e,\mu_R^2,s) C_F ~\Bigg[~ 
       \Upsilon \left({\e} \right)
           {\Msq}_{q \qb, int}
      +        {\kappa^2 {\cal R}e {\cal D}(s)}{e_q^2 \over 2 N}  ~
                   \Big\{   (17 -8\zeta(2))t^2 
\nonumber\\[1ex]
&&                         -(2tu + 3u^2) \ln\frac{-t}{s} 
                         - \left(2tu +2t^2 +u^2 \right) \ln^2\frac{-t}{s}
                   +t \leftrightarrow u \Big\}  
\nonumber\\[1ex]
&&          -{\kappa^2 \pi {\cal I}m {\cal D}(s)}{e_q^2 \over 2 N}  ~
                   \Big\{   3 t^2 + 2 t u
                         +2 (t^2 +2 t u + 2 u^2) \ln\frac{-u}{s}
                   +t \leftrightarrow u \Big\}
 \Bigg] 
\\[2ex]
\overline{|M^{V}|^2}_{gg,int}
  &=&
       a_s(\mu_R^2)~{e_q^2 \kappa^2}~{1 \over N^2-1}  ~\Bigg[~
              s~{\cal R}e {\cal D}(s) ~ \Big\{   u^2
                         +(2tu + t^2) \ln\frac{-u}{s}
\nonumber \\[1ex]
&&                         + \left(u^2+{1 \over 2}  t^2 + t u \right) \ln^2\frac{-u}{s}\Big\}
              +s~\pi{\cal I}m {\cal D}(s) ~ \Big\{   u^2 +2 t u
\nonumber\\[1ex]
&&                         +(2 u^2 +2 t u + t^2 ) \ln\frac{-u}{s} \Big\}
                   +t \leftrightarrow u    \Bigg]\,,
\eea
and the pure gravity contributions are
\bea
\overline{|M^{V}|^2}_{q \qb,gr} 
  &=&
       a_s(\mu_R^2) f(\e,\mu_R^2,s) C_F ~\Bigg[~ 
       \Upsilon \left({\e} \right)\Msq_{q\qb, gr}
      +  4(2\zeta(2) -5)  \Msq_{q\qb, gr}
       ~ \Bigg]  
\\ [2ex]
\overline{|M^{V}|^2}_{gg,gr} 
  &=&
       a_s(\mu_R^2) f(\e,\mu_R^2,s) C_A ~\Bigg[~ 
 \left\{-\frac{16}{\e^2} + \frac{4}{C_A \e}  \left( {11 \over 3 } C_A -\frac{4}{3}n_f T_f\right)\right\}  
{\Msq}_{gg, gr}
\nonumber \\[1ex]
              &+& \frac{1}{9} \left( 72 \zeta(2) + 70 \frac{n_f T_f}{C_A} -203  \right) 
{\Msq}_{gg, gr}
\Bigg] 
\eea
where
\bea
\Upsilon \left({\e} \right)
         &=& -~\frac{16}{\displaystyle{\e^2}} + \frac{12}{\displaystyle{\e}}  ,\quad \quad \quad
f(\e,\mu_R^2,s)  = \frac{ \Gamma\left( 1+{\displaystyle {\e\over 2}}\right)} {\Gamma(1+\e)} 
             \left( \frac{s}{4 \pi \mu_R^2} \right)^{\frac{\displaystyle{\e}}{2}}
\eea 
Here $\mu_R^2$ is the scale at which the theory is 
renormalized ; $a_s(\mu_R^2) =g_s(\mu_R^2)^2/16 \pi^2$
is the strong running coupling constant.
Our results for the SM are in agreement with the literature \cite{Bailey:1992br}.
The poles in $\epsilon$ arise from loop integrals and correspond
to the soft and collinear divergences. Configurations in which
a virtual gluon momentum goes to zero give soft singularities while 
collinear singularities arise when two massless partons become 
collinear to each other. 
As the soft divergences cancel completely in any observable, 
the $\e$ poles of order-2, get canceled when real
emission contributions are included. This cancellation will be 
shown in what follows.   
\subsection {Real emission process}   
A next-to-leading order $2 \rarrow 3$ parton level process
for production of photon pairs is of the following generic
form 
\be
a(p_1) + b(p_2) \rarrow \gamma(p_3) + \gamma(p_4) + c(p_5).
\ee
where a,b and c are massless partons. In fig.(\ref{real}) all gravity
mediated $2 \rarrow 3$ Feynman diagrams are given. Depending
on the initial state partons, the final state may 
have a quark or anti-quark or a gluon. To obtain an 
inclusive cross-section the final state parton will be
integrated over the phase-space. The $2 \rarrow 3$ matrix
elements when integrated over the phase-space give soft
and collinear singularities. These singularities are 
regulated using dimensional regularization with $n=4+\e$ 
and appear as poles in $\e$. These singularities arise when
the final state gluon becomes soft (a soft fermion does not
give any soft divergences) or when the final state massless
parton becomes collinear to an initial state massless parton.
As was mentioned in the introduction a Monte Carlo approach
allows for an easy implementation of experimental cuts on
the final state photons and smooth-cone isolation criterion.
This is achieved by using the semi-numerical 
two cutoff phase space slicing method. This method 
introduces two small dimensionless parameters $\delta_s$ and $\delta_c$ 
to deal with soft and collinear QCD singularities.  $\delta_s$  
divides the phase-space into {\it soft} and {\it hard} regions.
The part of phase-space where the energy of the gluon in the
centre of mass frame of incoming parton is less than 
$\delta_s \sqrt{ s}/2$ is defined as ${\it soft}$ and the
region complementary to it is ${\it hard}$. For small values
of $\delta_s$ the matrix elements can be simplified and 
integrated over the soft region to give a $\delta_s$ dependent,
order $\alpha_s$, 2-body contribution $d\sigma_S(\delta_s, \e)$.
This contains the poles in $\e$ arising from the soft singularities.
The hard region can be further divided into collinear and 
non-collinear regions using another small dimensionless slicing
parameter $\delta_c$. The part of phase-space in which the 
final state parton is collinear to the incoming parton is 
defined as collinear region and gives an order $\alpha_s$
contribution $d\sigma_{HC} (\delta_s, \delta_c, \e)$. This 
contains the collinear singularities. The hard non-collinear 
3-body contribution denoted by $d\sigma_{\overline {HC}} (\delta_s, \delta_s)$
is free of any singularities and can be evaluated numerically
using Monte Carlo integration. The collinear singularities
appearing in $d\sigma_{HC}(\delta_s, \delta_c, \e)$ will be removed
by mass factorization in ${\overline{MS}}$ scheme by adding
counter terms to give $d \sigma_{HC + CT}(\delta_s, \delta_c, \e)$.
In the following subsections it will be shown that the 2-body
contribution 
$
 d\sigma_V (\e) 
+ d\sigma_S (\delta_s,\delta_c, \e) 
+ d\sigma_{HC+CT} (\delta_s,\delta_c, \e)
$
is free of poles in $\e$. Although individually the 2-body and 3-body 
contributions depend on the slicing
parameters which were introduced artificially in the problem, the sum
should be independent of these parameters.
In what follows we will show that this sum is independent of $\delta_s$
and $\delta_c$ for a fairly wide rage of these parameters. 
\subsubsection{Soft}
In the soft gluon limit the $2 \rarrow 3$ amplitude factorizes into 
Born matrix element and a term containing eikonal currents. These 
eikonal currents reveal the singularities when integrated over the
soft part of phase space.
\bea
{d { \hat \sigma}}_S = a_s(\mu_R^2) f(\e,\mu_R^2,s) 
\Big( C_F~ {d \hat \sigma^0_{q \qb}}(\e)  + C_A~ {d \hat \sigma^0_{gg}}(\e) \Big )
\left[ \frac{16}{\e^2} + \frac{16}{\e} \ln \delta_s + 8 \ln^2 \delta_s \right]
\eea
The symbol ${\hat \sigma}$ is used to indicate that the cross-section is at parton level.
The terms linear and higher order in $\delta_s$ have been dropped. Note 
that the $1/\e^2$ pole cancels with the virtual contribution. However the
pole $1/\e$ with coefficient $\ln \delta_s$ still remains uncanceled and
later it will be seen that this pole also cancels.
\subsubsection{Collinear}
Complementary to the soft region discussed above is 
the hard region. In this region collinear singularities
arise when the final state massless parton (quark, anti-quark or gluon)
is collinear to the initial state parton. 
Let $z$ denote the momentum fraction of the incoming parton
carried by the parton entering into hard scattering. An initial
state quark can split into a quark (and a gluon) or into a gluon
(and a quark) which enter into the hard scattering and involve
$P_{qq}$ and $P_{gq}$ splitting functions. Similarly an initial
state gluon gives $P_{gg}$ and $P_{qg}$ splitting functions.
If the energy of a final state gluon is greater than 
$\delta_s \sqrt{s} /2 $ in the rest frame of incoming
partons it is defined as a hard gluon. Thus a gluon is hard if
$0\leq z \leq 1-\delta_s$ for $P_{qq}$ and $P_{gg} $ splittings.
As a soft quark does not give any soft singularities,
$0 \leq z \leq 1$ for $P_{gq}$ and $P_{gq}$ splittings.
As already discussed above,
in the collinear limit matrix elements simplify and can be integrated
easily in $n=4+\e$ dimensions over the collinear region. For
photon pair production the hard collinear contribution takes the following form
\bea
 d \sigma_{HC}\!\! &=& \!\!\frac{a_s(\mu_R^2)}{\e} dx_1 dx_2 f(\e, \mu_R^2, s)  
\nonumber \\[2ex] 
&& \times
\Bigg[~   d {\hat \sigma}_0^{q\qb}(x_1,x_2,\e)   
             \Big\{  \ixttomds P_{qq}(z,\e) \sum_i f_{q_i}(x_1) f_{\qb_i}(x_2/z) 
\nonumber \\[2ex]
&&                 + \ixotomds P_{qq}(z,\e) \sum_i f_{q_i}(x_1/z) f_{\qb_i}(x_2)  
                   + x_1 \leftrightarrow x_2 
              \Big\}_{~ q \qb~ } 
\nonumber \\ [2ex]
&&      + d {\hat \sigma}_0^{q\qb}(x_1,x_2,\e) 
             \Big\{  \ixtto P_{qg}(z,\e) \sum_i f_{q_i}(x_1) f_{g}(x_2/z) 
\nonumber \\ [2ex]
&&                   + \ixtto P_{qg}(z,\e) \sum_i f_{\qb_i}(x_1) f_{g}(x_2/z)
                   + x_1 \leftrightarrow x_2
              \Big\}_{~ qg ~ } 
\nonumber \\ [2ex]
&&      + d {\hat \sigma}_0^{gg}(x_1,x_2,\e) 
             \Big\{  \ixttomds P_{gg}(z,\e) \sum_i f_{g}(x_1) f_{g}(x_2/z)
                   + x_1 \leftrightarrow x_2
             \Big\}_{~ gg ~ }
\nonumber \\ [2ex]
&&      + d {\hat \sigma}_0^{gg}(x_1,x_2,\e) 
             \Big\{  \ixttomds P_{gq}(z,\e) \sum_i f_{g}(x_1) f_{q_i}(x_2/z) 
\nonumber \\ [2ex]
&&                   + \ixttomds P_{gq}(z,\e) \sum_i f_{g}(x_1) f_{\qb_i}(x_2/z)
                   + x_1 \leftrightarrow x_2
             \Big\}_{~ qg ~ } 
\Bigg] 
\eea
where $x_1,x_2$ are momentum fraction of incoming parton momenta
$P_{ij}(z,\e) $ are splitting functions in $4+\e$ dimensions, and
\be
{\cal H}(z,\e,\delta_c) = \left( \delta_c  \frac{1-z}{z} \right)^{\e/2}.
\ee

The collinear singularities can be removed by the method of mass factorization. To this
effect, counter terms to cancel these singularities in $ {\overline {MS}} $ scheme are obtained 
by introducing in the leading order cross-section
\bea
d \sigma_0 &=& dx_1 dx_2 
   \Big(  d {\hat \sigma}_0^{q\qb}(x_1,x_2,\e)  \sum_i \Big[ 
f_{q_i}(x_1) f_{\qb_i}(x_2)   +f_{\qb_i}(x_1) f_{q_i}(x_2) \Big]  
\nonumber\\[2ex]
&&          + d {\hat \sigma}_0^{gg}(x_1,x_2,\e) f_{g}(x_1)f_{g}(x_2) 
  \Big) 
\eea
the following factorization scale dependent parton distribution functions in 
the $ {\overline{MS}}$ scheme.
\bea
f_q(x)\!\!\! &=&\!\!\! f_q(x, \mu_F) - \frac{a_s(\mu_R^2)}{\e} \left( \frac{\mu_F^2}{\mu_R^2} 
\right)^{\displaystyle{\e\over 2}}
      \int_x^1 \frac{dz}{z} \Big[ P_{qq}(z)f_q(x/z) + P_{qg}(z)f_g(x/z) \Big]   
\nonumber \\[2ex]
f_g(x)\!\!\! &=&\!\!\! f_g(x, \mu_F) - \frac{a_s(\mu_R^2)}{\e} \left( \frac{\mu_F^2}{\mu_R^2} 
\right)^{\displaystyle{\e\over 2}}  
       \int_x^1 \frac{dz}{z} \Big[ P_{gg}(z)f_g(x/z) + P_{gq}(z)\big(f_q(x/z) 
                    + f_{\qb}(x/z)\big)  \Big]
\nonumber\\
\eea        
%
Note that the upper limits on the integrals are
1 for all the splittings. Substituting these distribution
functions in $d\sigma_0$ an adding to $\sigma_{HC}$ the following
order $a_s$ term is obtained.
\newcommand{\epsln}{\ensuremath{ \Big(-\frac{1}{\e}+ \frac{1}{2} \ln \frac{s}{\mu_F^2}  \Big)}} 
\bea
d \sigma_{HC + CT}\!\!\!  &=& \!\!\!  a_s(\mu_R^2) dx_1 dx_2 f(\e, \mu_R^2, s) 
\Bigg[ d{\sigma_0}^{q\qb}(\e)\sum_i f_{\qb_i}(x_1,\mu_F) 
\Bigg\{\frac{1}{2} {\tilde f}_{q_i}(x_2,\mu_F)
\nonumber\\[2ex]
&&             +\Big( f_{q_i}(x_2,\mu_F) A_{q \rarrow q+g}
             +   f_{g}(x_2,\mu_F) A_{g \rarrow q+\qb} \Big)  
\epsln   \Bigg \}
\nonumber\\[2ex]
&&+ d{\sigma_0}^{gg}(\e) f_{g}(x_1,\mu_F) 
\Bigg\{\frac{1}{2} {\tilde f}_{g}(x_2,\mu_F)
             +\Big( f_{g}(x_2,\mu_F) A_{g \rarrow g+g}
\nonumber\\[2ex]
&&             +   \sum_i f_{q_i}(x_2,\mu_F) A_{q \rarrow g+q} \Big)  
\epsln   \Bigg \} \Bigg] + (q \leftrightarrow \qb,x_1 \leftrightarrow x_2)  
\eea
The function $A_{a \rarrow b+c}$ result from the mismatch in the 
integral limits on $z-$integrals in $d\sigma_{HC}$ and counter term
and can be easily evaluated using the definition of {\it plus}-prescription.
\bea
A_{q \rarrow q +g}   &=& 4 C_F \left(2 \ln \delta_s + \frac{3}{2} \right),   
\quad \quad \quad \quad \quad \!\!\!
A_{q \rarrow g +q}   = 0.
\nonumber \\
A_{g \rarrow g + g}  &=& \frac{22}{3}C_A - \frac{4}{3}n_f +8C_A \ln \delta_s,  
\quad \quad
A_{g \rarrow q +\qb} = 0,
\eea
The function ${\tilde f_{q,g}}$ are defined by
\bea
{\tilde f}_{q}(x, \mu_F)
    &=&  \int_x^{1 -\delta_s} \frac{dz}{z} f_q \left( \frac{x}{z}, \mu_F \right) {\tilde P}_{qq}(z)
      + \int_x^{1}            \frac{dz}{z} f_g \left( \frac{x}{z}, \mu_F \right) {\tilde P}_{qg}(z)
\nonumber \\[2ex]
{\tilde f}_{g}(x, \mu_F)
    &=&  \int_x^{1 -\delta_s} \frac{dz}{z} f_q \left( \frac{x}{z}, \mu_F \right) {\tilde P}_{gq}(z)
      + \int_x^{1}            \frac{dz}{z} f_g \left( \frac{x}{z}, \mu_F \right) {\tilde P}_{gg}(z)
\eea
with
\be
{\tilde P}_{ij}(z) = P_{ij}(z) \ln \left( \delta_c \frac{1-z}{z} \frac{s}{\mu_F^2} \right) - P^{\prime}_{ij}(z)
\ee
The $ P^{\prime}$ is the order $\e$ part of $P_{ij}(z,\e)$.

After mass factorization the poles still remain and do not cancel completely in
$d\sigma_{HC + CT}$ and these cancel with the uncanceled simple poles in virtual
and soft contributions. The contribution $d\sigma_{HC +CT} + d\sigma_{S} +d\sigma_{V}$
is an order $a_s$ 2-body contribution free of any singularities and can be evaluated
numerically using Monte Carlo integration with the experimental cuts on the final state photons.
This, however 
depends on the choice of  arbitrary small parameters $\delta_s$ and $\delta_c$ used for slicing
of phase space. The 3-body hard non collinear contribution also depends on slicing parameters
and is free of any singularities and can be evaluated numerically. The sum of 2-body and 3-body
contribution should be independent of the $\delta_s$ and $\delta_c$.

\section{Numerical Results}
In this section various kinematical distributions for production 
of isolated direct photon pairs are presented to next-to-leading 
order accuracy in QCD both in the ADD and RS scenarios. Both
for the SM background and the SM+ADD and SM+RS signals the following
distributions are presented:
\begin{enumerate}
\item Invariant mass ($Q$) distribution of the di-photon system
\item Transverse momentum ($Q_T$) distribution of the photon pair 
\item Angular distribution ${ \cos}\theta^*$ of the photons
\item Rapidity $(Y)$ distribution of the di-photon system. 
\item Rapidity $y^\gamma$ distribution of photon
\end{enumerate}
We impose the same kinematical cuts on the two photons in our
study which are used by ATLAS and CMS collaborations \cite{atlas,cms}:
(i) $p_T^\gamma >~40~(25)$ GeV for the harder (softer) photons,
(ii) rapidity $|y_\gamma| < 2.5$ for each of the photons.
(iii) The minimum separation between the two photons in the $y-\phi$ plane
is taken to be $R_{\gamma\gamma}~=~0.4$. 
As this study does not include poorly known fragmentation functions, the 
final state QED singularity is suppressed using the smooth cone isolation
discussed in eqn.(\ref{HR}).
In what follows $E_T^{iso}=15$ GeV and $n=2$ with $R_0=0.4$ would be 
our default choices. This choice allows a maximum of hadronic transverse
energy equal to 15 GeV in a cone of radius 0.4 in $\eta-\phi$ plane around
a photon.  As parton in the final state in our NLO computation
is a crude approximation to the jet of hadrons detected in the detectors,
the dependence of results on the choice of the parameters entering
into isolation criterion needs to be studied and it has been observed 
in \cite{ourpapers2} to be small.
For our leading order analysis we have used CTEQ6L, and for NLO analysis CTEQ6M 
\cite{Pumplin:2002vw} parton density sets respectively with $n_f = 5$
light quark flavours, and the corresponding two loop strong running coupling constant 
$\alpha_s (M_Z) = 0.118$. The fine structure constant is taken 
to be $\alpha(M_W)=1/128$.  
Unless mentioned otherwise we have 
set the renormalization and the factorization scales to 
$\mu_R = \mu_F = Q$ in all the distributions.

\begin{figure}[ht]
\centerline{\epsfig{file=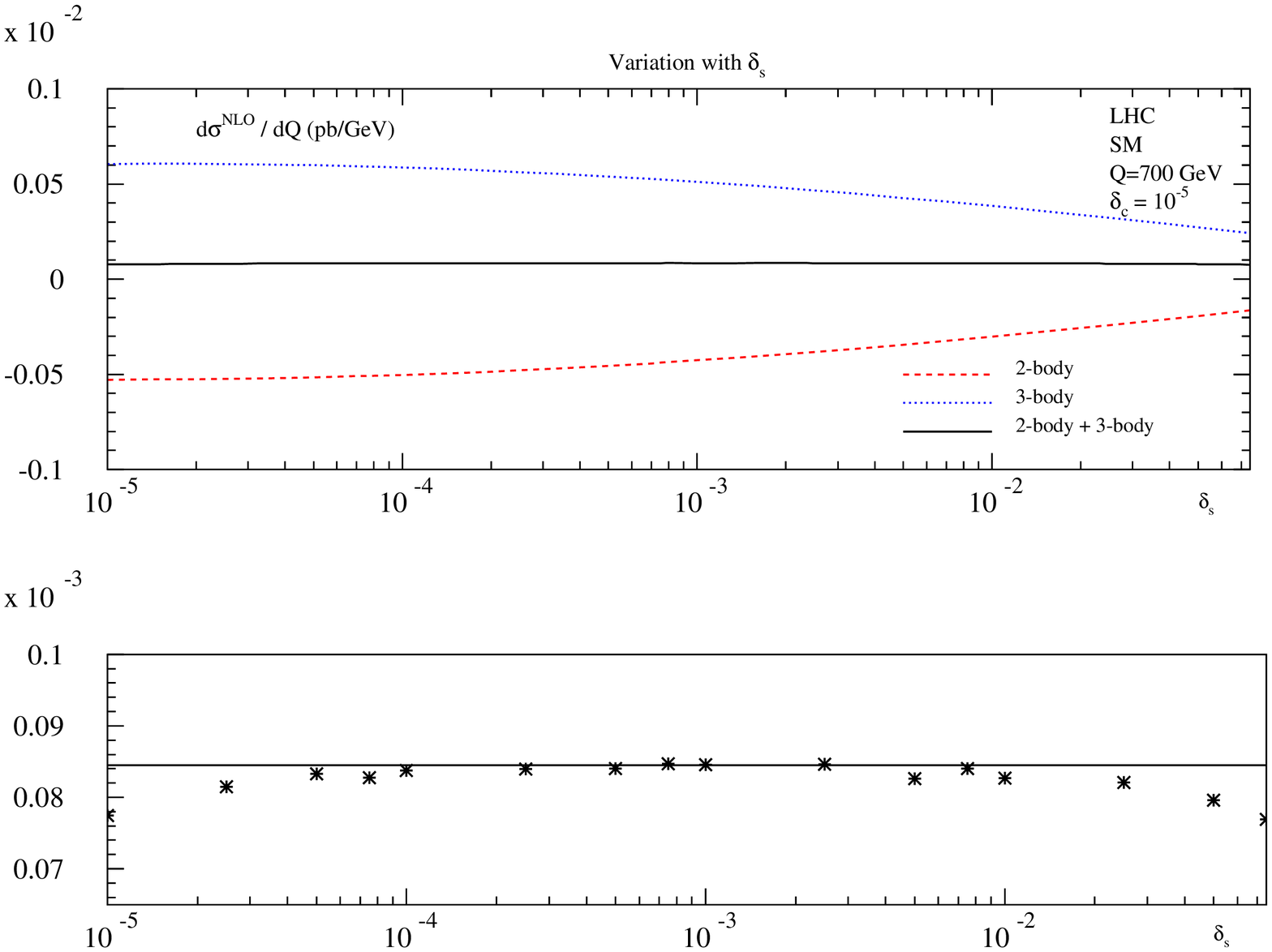,width=15cm,height=12cm,angle=0}}
\caption{Stability of the order $\alpha_s$ contribution
to the SM cross section against the variation
of the slicing parameter $\delta_s$ (top), with $\delta_c = 10^{-5}$ fixed,
in the invariant mass distribution
of the di-photon.  Below is shown the variation of the sum of $2$-body and $3$-body
contributions over the range of $\delta_s$ considered and contrasted 
against the one at $\delta_s = 10^{-3}$.}
\label{sm-add-deltas}
\end{figure}

\begin{figure}[htb]
\centerline{\epsfig{file=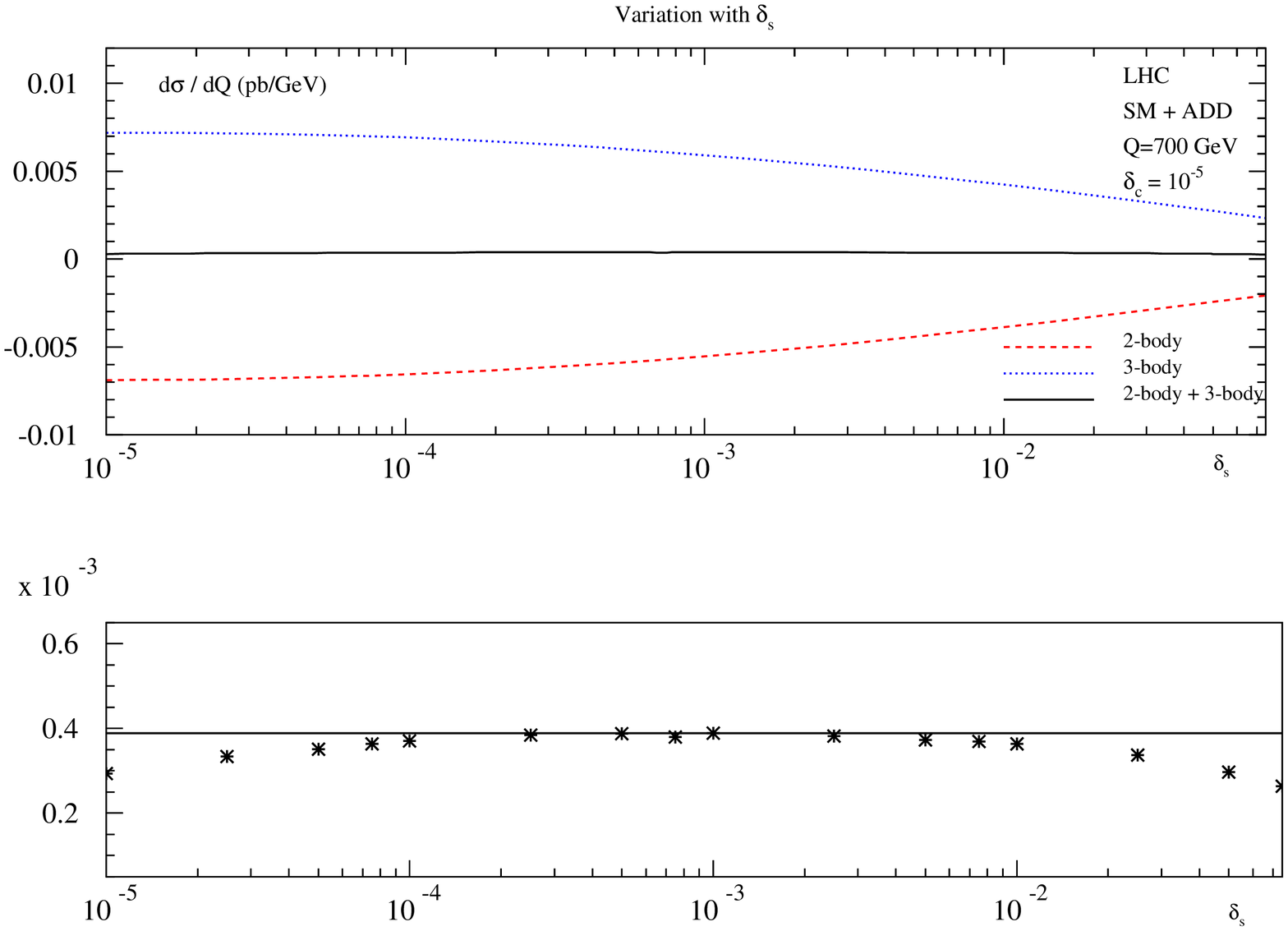,width=15cm,height=12cm,angle=0}}
\caption{Stability of the order $\alpha_s$ contribution to the SM+ADD cross 
section against the variation of the slicing parameter $\delta_s$ (top), 
with $\delta_c = 10^{-5}$ fixed, 
in the invariant mass distribution of the di-photon with $M_S=2$ TeV and $d=3.$
Below is shown the variation of the sum of $2$-body and $3$-body
contributions over the range of $\delta_s$ considered and contrasted 
against the one at $\delta_s = 10^{-3}$.}
\label{add-deltas}
\end{figure}

\begin{figure}[htb]
\centerline{\epsfig{file=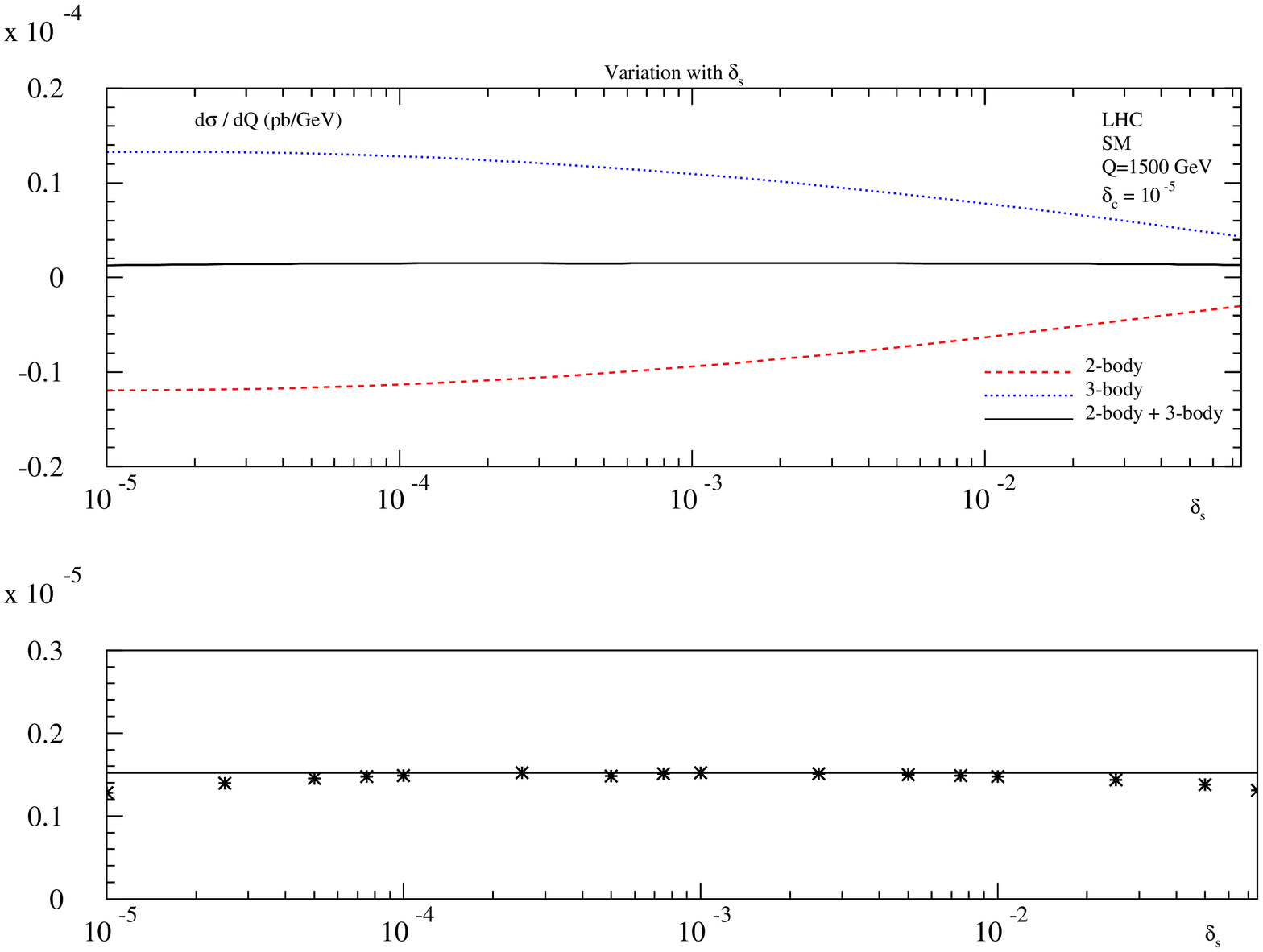,width=15cm,height=12cm,angle=0}}
\caption{Stability of the order $\alpha_s$ contribution
to the SM cross section against the variation
of the slicing parameter $\delta_s$ (top),  with $\delta_c = 10^{-5}$ fixed, 
in the invariant mass distribution
of the di-photon.  Below is shown the variation of the sum of $2$-body and $3$-body 
contributions over the range of $\delta_s$ considered and contrasted 
against the one at $\delta_s = 10^{-3}$.}
\label{sm-rs-deltas}
\end{figure}

Before proceeding further we present the stability of the sum of
$2$-body and $3$-body contributions against the variation of the 
slicing parameters $\delta_s$ and $\delta_c$. In fig(\ref{sm-add-deltas}) and fig(\ref{add-deltas})
the individual 2-body and 3-body order $\alpha_s$ contributions and 
their sum are presented in invariant mass distribution in the SM
and SM+ADD respectively as a function of $\delta_s$ with $\delta_c$ fixed
at $10^{-5}$. From these figures it is clear that the sum is fairly
stable against the variation of slicing parameters; this serves as a 
check on the numerical implementation of the phase space slicing in our
numerical code. Fig(\ref{sm-rs-deltas}) provides the corresponding test for the case
of the RS model.
For all further analysis, we choose $\delta_s = 10^{-3}$ and $\delta_c = 10^{-5}$.
As a further test we compared our SM results with those
in \cite{Bern:2002jx}.
Our SM results are in good agreement with \cite{Bern:2002jx} when the isolation
criterion used there is used in our code. This gives a further 
confidence in our code.

\subsection{ADD model distributions}
\begin{figure}[htb]
\centerline{
\epsfig{file=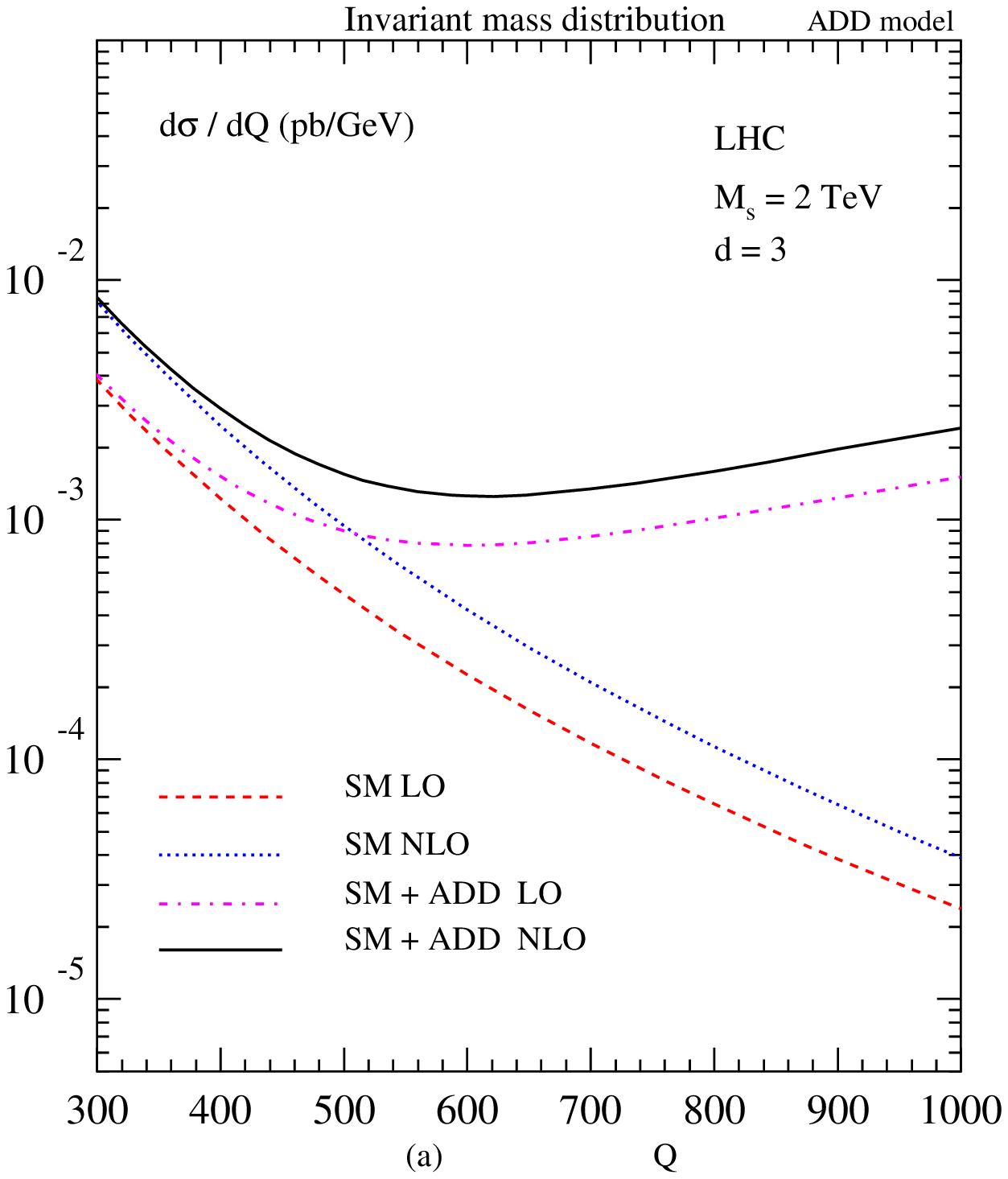,width=8.5cm,height=9.5cm,angle=0}
\epsfig{file=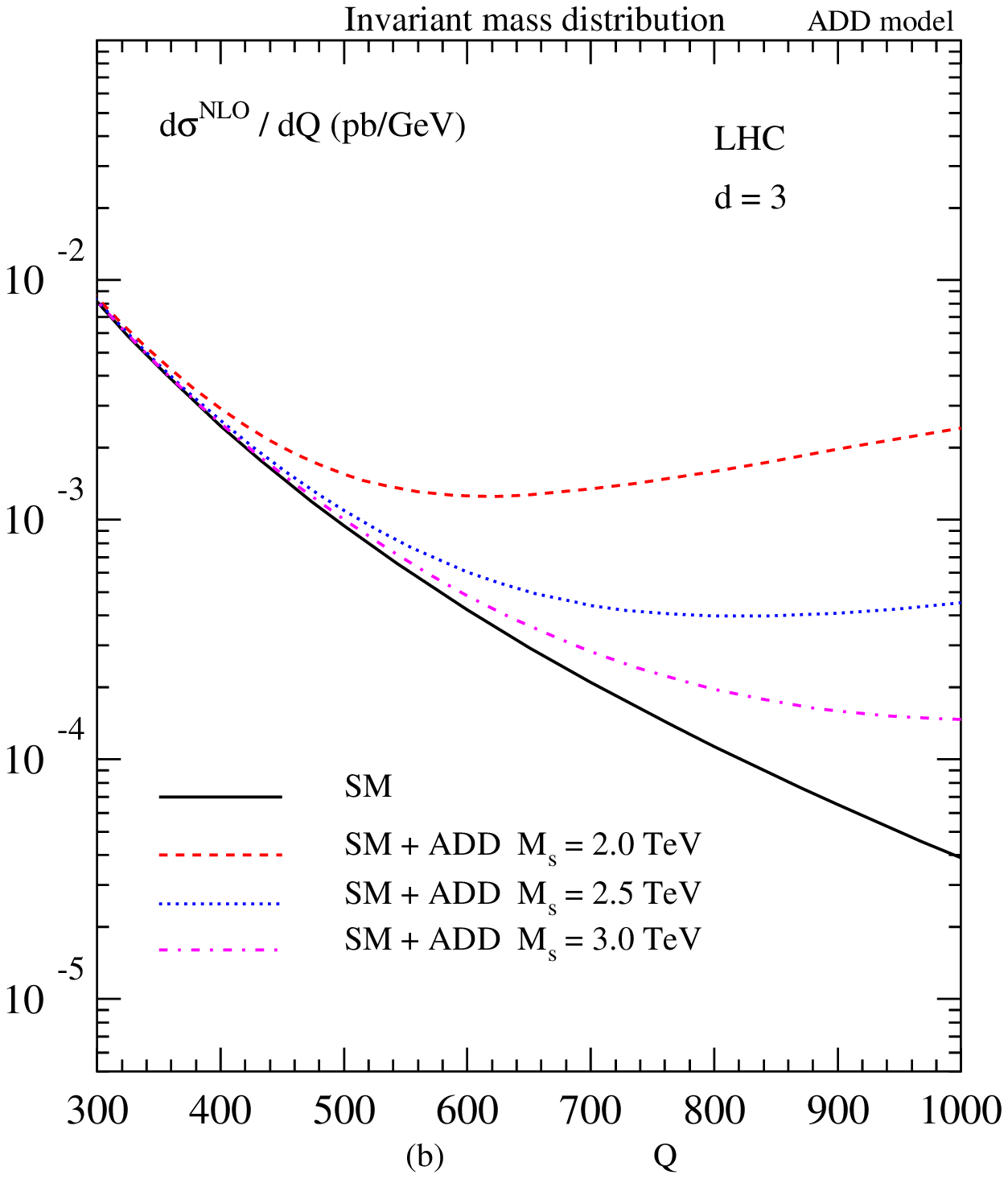,width=8.5cm,height=9.5cm,angle=0}}
\centerline{\epsfig{file=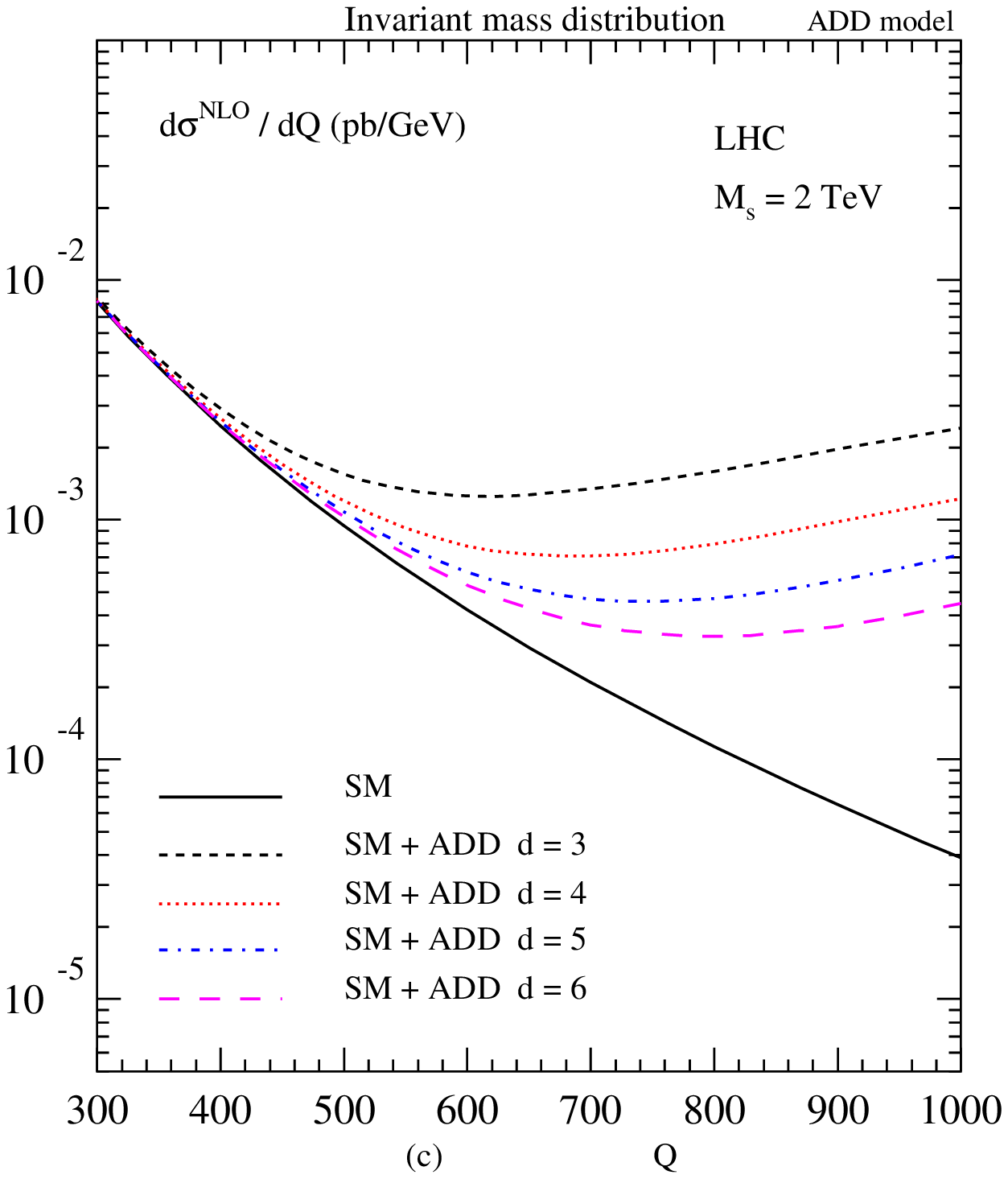,width=8.5cm,height=9.5cm,angle=0}
\epsfig{file=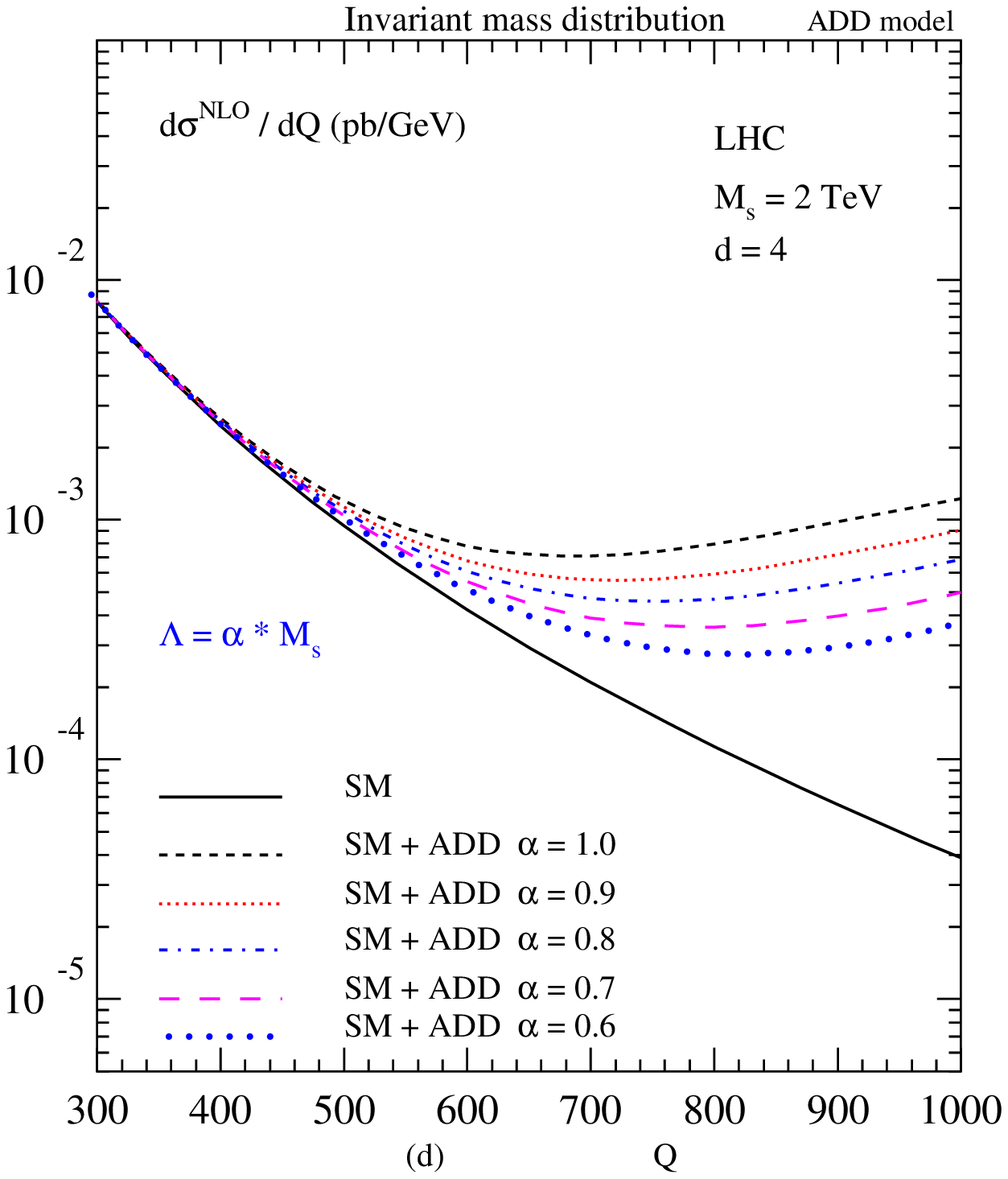,width=8.5cm,height=9.5cm,angle=0}}
\caption{Invariant mass distribution of the di-photon production in the ADD
model at the LHC.  
In (a) both SM and the signal (SM+ADD) are presented at LO
and NLO for $M_S=2$ TeV and $d=3$.  
Further the dependency of the cross sections on 
the scale $M_S$ in (b), on the number $d$ of extra dimensions in (c) and on the  
cut-off scale $\Lambda$ for the summation over virtual KK modes in (d), has been
shown to NLO in QCD.}
\label{q-dist}
\end{figure}

In this section, we study various kinematic distributions
in the ADD model using our NLO results.  First, we present our results for the invariant mass distribution. 
In fig.(\ref{q-dist}a), we plot LO and NLO contributions to 
the signal (SM+ADD) and the SM background against $Q$ between $300~$GeV and
$1~$TeV.
We choose the fundamental scale $M_S=2$ TeV 
and the number of extra dimensions $d=3$.  
As we discussed in the introduction, we do not consider the gluon-gluon fusion process through
quark loop at LO as its contribution is 
significant only at small $Q$. 

For the above choice of parameters the signal starts deviating from the
SM background around $Q=500$ GeV.
The value of $Q$ at which the deviation occurs depends very much on the choice of the parameters, 
namely the scale $M_S$, $d$ and the cut-off scale $\Lambda$ for the summation of the KK modes.
In fig.(\ref{q-dist}b), we show how the invariant mass distribution depends
on the choice of the fundamental scale $M_S$ when $d=3$.  As expected
smaller the $M_S$, the larger the deviation one observes.
The dependence on the number of extra dimensions $d$ is presented in 
fig.{\ref{q-dist}c} for $d=3-6$ keeping $M_S=2$ TeV fixed.  
We find that the ADD contribution decreases with increase in $d$.
In fig.(\ref{q-dist}d), we present
the cut-off scale $\Lambda$ dependence for $\Lambda=0.6M_S$ to $M_S$.
For lower values of cut-off scale, the number of KK modes available are less
and the signal will decrease with decrease in $\Lambda$ as shown in the 
figure.  In the following, we choose $M_S = 2$ TeV, $d = 3$ and $\Lambda = M_S$.
For the rest of the kinematic distributions that we have considered, 
to reduce the SM background and to enhance the signal, we integrate over $Q$ in the range
$600 < Q < 1100$ GeV. 

\begin{figure}[htb]
\centerline{
\epsfig{file=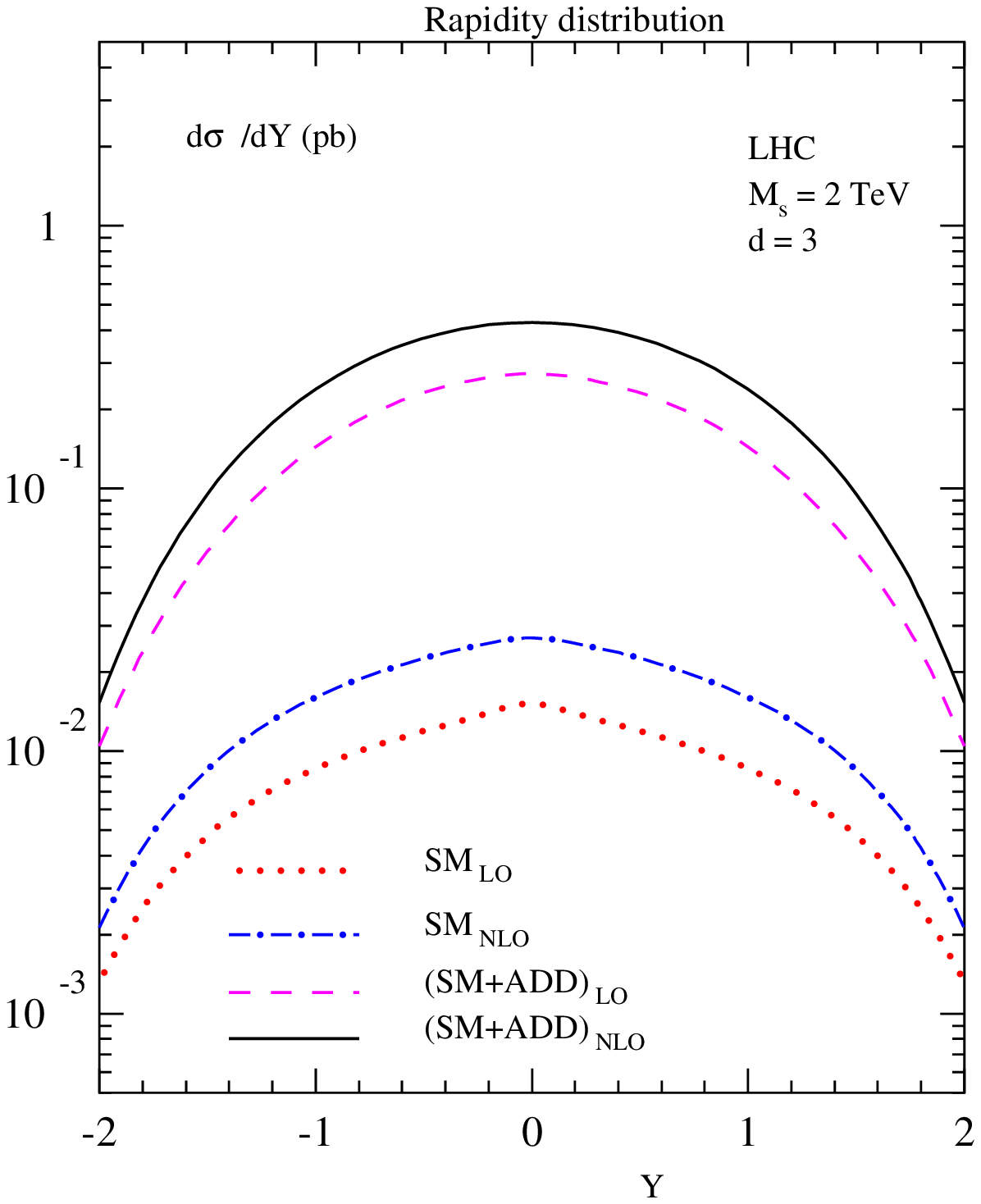,width=7.5cm,height=8.5cm,angle=0}
\epsfig{file=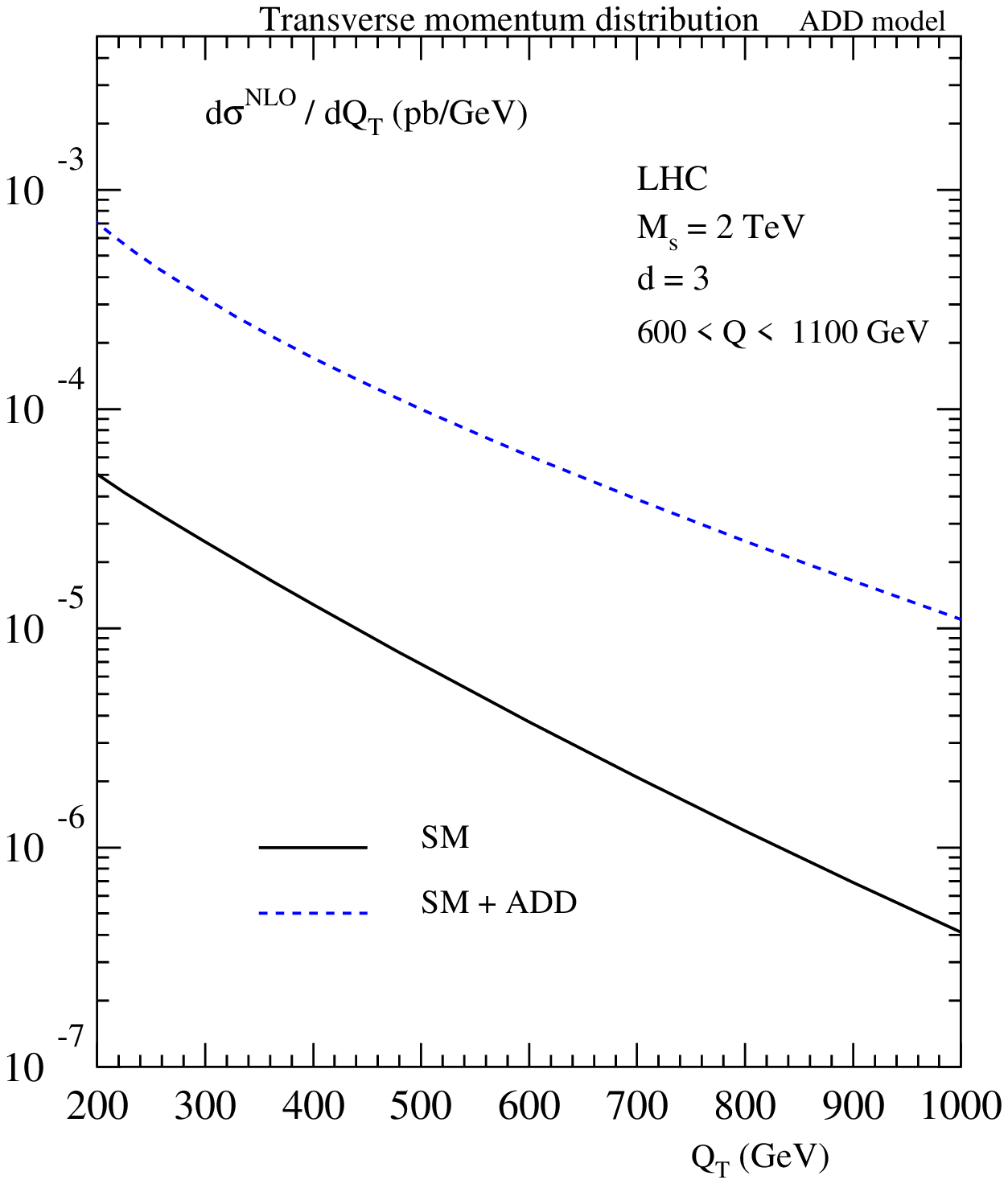,width=7.5cm,height=8.5cm,angle=0}}
\caption{Transverse momentum rapidity $d\sigma/dY$ (left) and 
$d\sigma/dQ_T$ (right) distributions of the di-photon production are 
presented in the ADD model with $M_S=2$ TeV, $d=3$ and by integrating 
over $Q$ in the range $600~<Q~<1100$ GeV.}
\label{add-yf-qt}
\end{figure}

The rapidity of the photon pair is defined by
\begin{eqnarray}
Y = \frac{1}{2}~{\ln}\left( \frac{P_1.q}{P_2.q}\right)
\label{eqn-Y}
\end{eqnarray}
where $P_1,P_2$ are the momenta of incoming hadrons and $q=p_3+p_4$. 
In the left panel of fig.(\ref{add-yf-qt}), we show the production cross section as a function of Y
between $-2.0$ and $2.0$ after integrating over $Q$ in the region 
$600 < Q < 1100$ GeV where the ADD model shows significant contribution 
over the SM background.
From the left panel of this figure, we observe that the signal exceeds the
background by more than an order of magnitude at the central rapidity
region $Y=0$.  

The transverse momentum of the photon pair is defined by $Q_T=\sqrt{q_x^2+q_y^2}$.  
At LO, the photon pairs will have zero $Q_T$ as incoming partons  
have no transverse momentum, and hence $Q_T$ distribution will be proportional to
$\delta(Q_T)$.  However, at NLO, the photon pairs will be accompanied by a quark (anti-quark)
or a gluon in the final state resulting in a non-zero $Q_T$.  The numerical results
for the $Q_T$ distribution is presented in the right panel of fig.(\ref{add-yf-qt}).

\begin{figure}[htb]
\centerline{
\epsfig{file=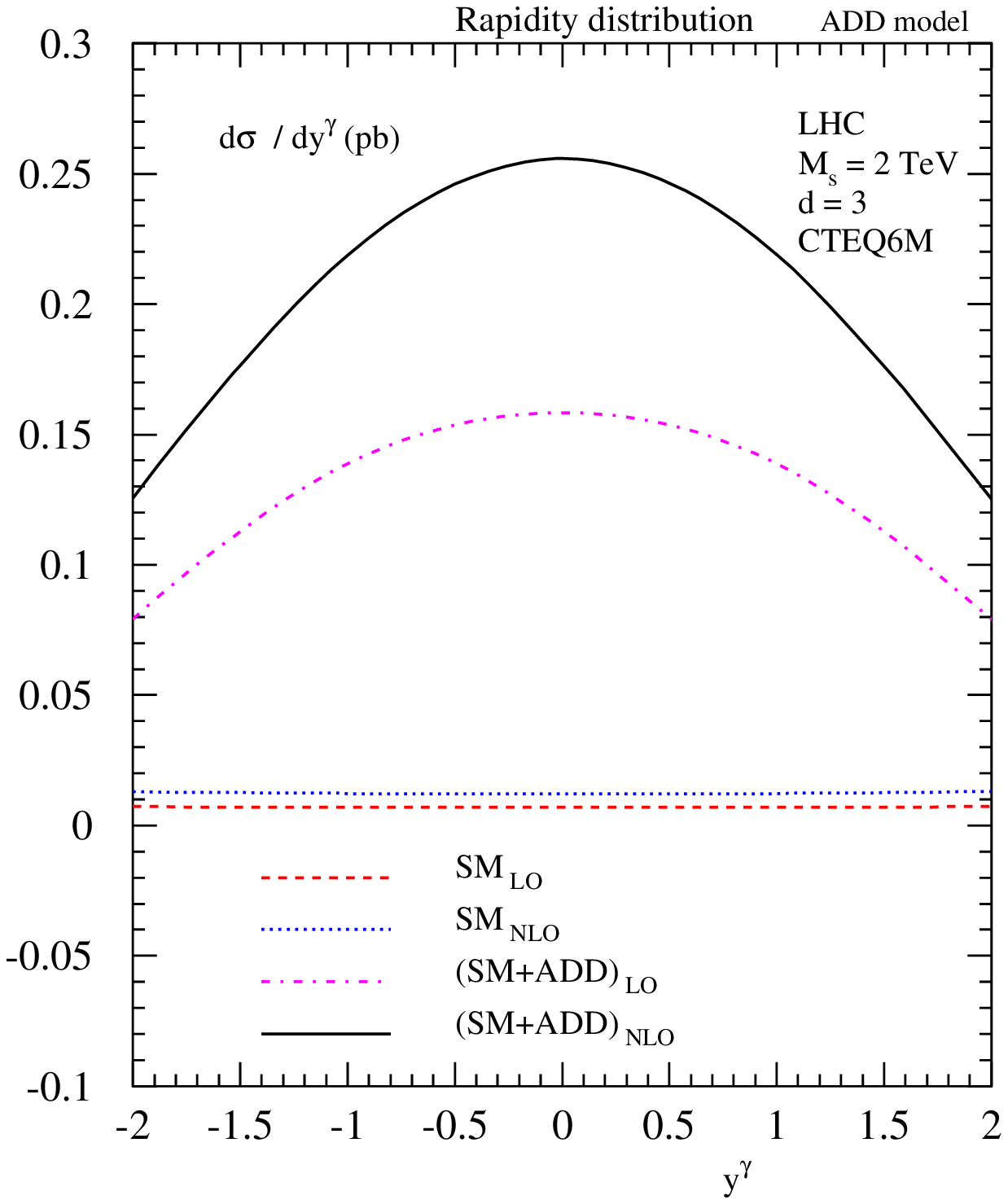,width=7.5cm,height=8.5cm,angle=0}
\epsfig{file=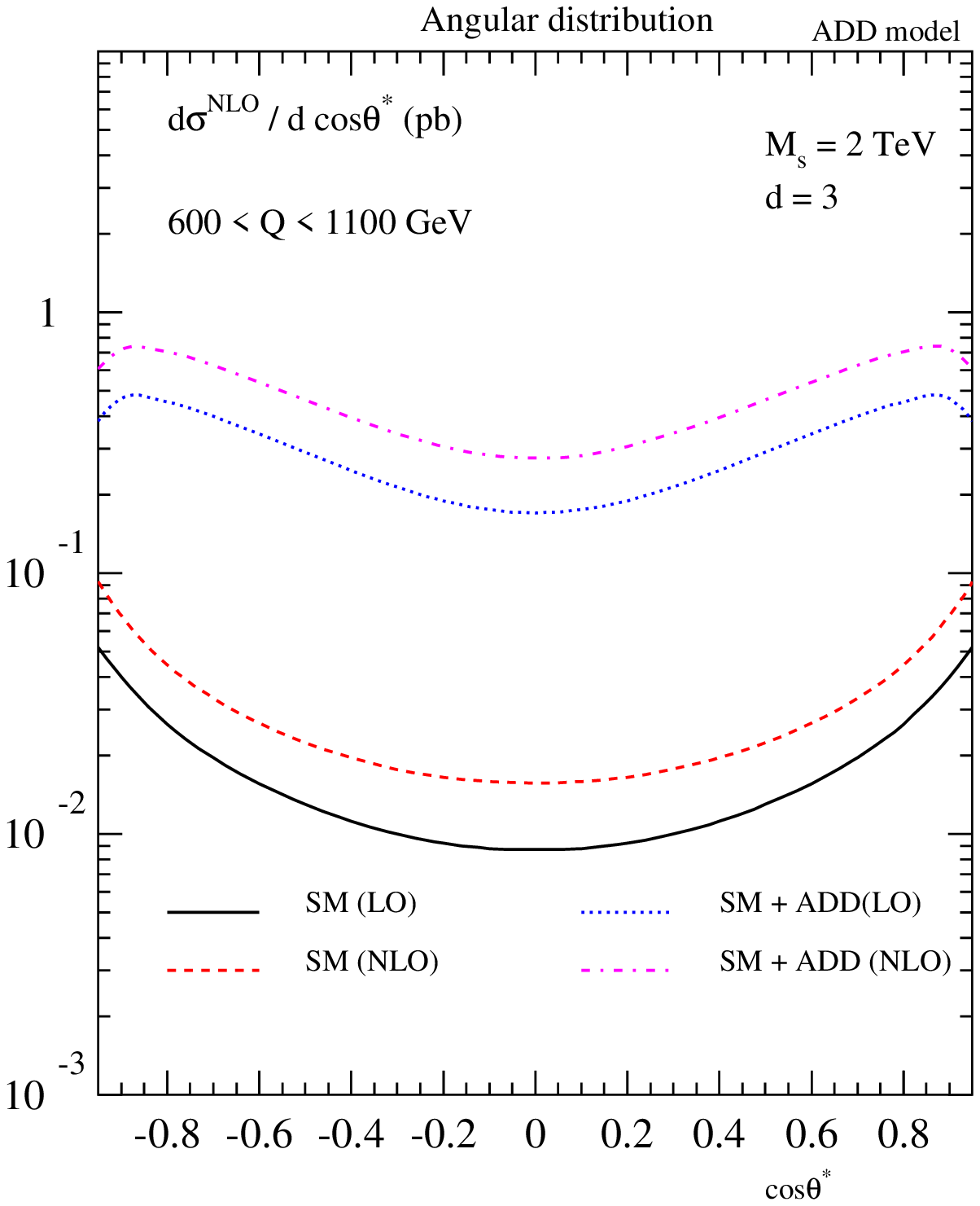,width=7.5cm,height=8.5cm,angle=0}}
\caption{Rapidity $d\sigma/dy^\gamma$ (left) and angular distributions 
$d\sigma/d~\cos\theta^*$ (right) of the photons are presented in the ADD model 
with $M_S=2$ and $d=3$.  Both of these distributions are obtained by 
integrating over the invariant mass of the di-photon in the range 
$600~<Q~<1100$ GeV.}
\label{y.ind-cf-dist}
\end{figure}

The rapidity of a photon is given by
\begin{eqnarray}
y^\gamma = \frac{1}{2}~~ \ln\left( \frac{E+p_z}{E-p_z}\right)
\label{eqn-y.ind}
\end{eqnarray}
where $E$ and $p_z$ are its energy and the longitudinal 
momentum respectively.  In fig.(\ref{y.ind-cf-dist}), the left panel shows the
rapidity distribution of the photons as a function of $y^\gamma$ in the region 
$-2.0 < y^{\gamma} < 2.0$.  The SM cross sections both at LO and NLO level
do not show significant dependence on $y^\gamma$ unlike  
contribution from the ADD model.  We also find that the QCD 
corrections are large for the signal as compared to the SM
background.

For the angular distribution of the photons, we define
\begin{eqnarray}
{\cos} \theta^* = \frac{P_1.(p_3-p_4)}{P_1.(p_3+p_4)}
\label{eqn-cf}
\end{eqnarray}
Since gravitons are spin-2 particles, the angular dependence of the
cross section in ADD model will be different from SM.  It is shown
in the right panel of fig.(\ref{y.ind-cf-dist}).
Hence these distributions have the advantage of 
distinguishing the signal, qualitatively, from the background. 
\subsection{RS model distributions}
\begin{figure}[htb]
\centerline{
\epsfig{file=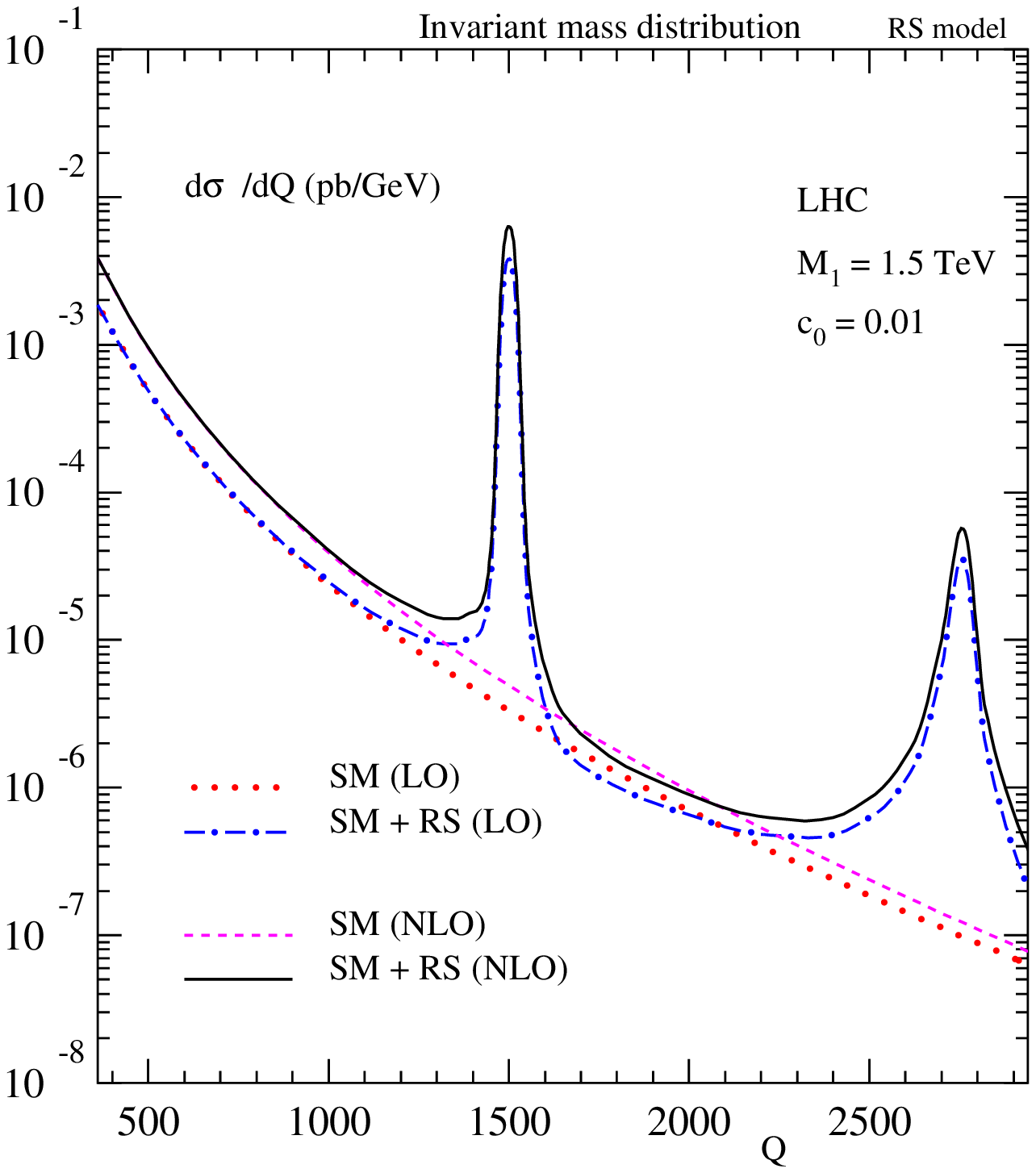,width=8cm,height=9cm,angle=0}
\epsfig{file=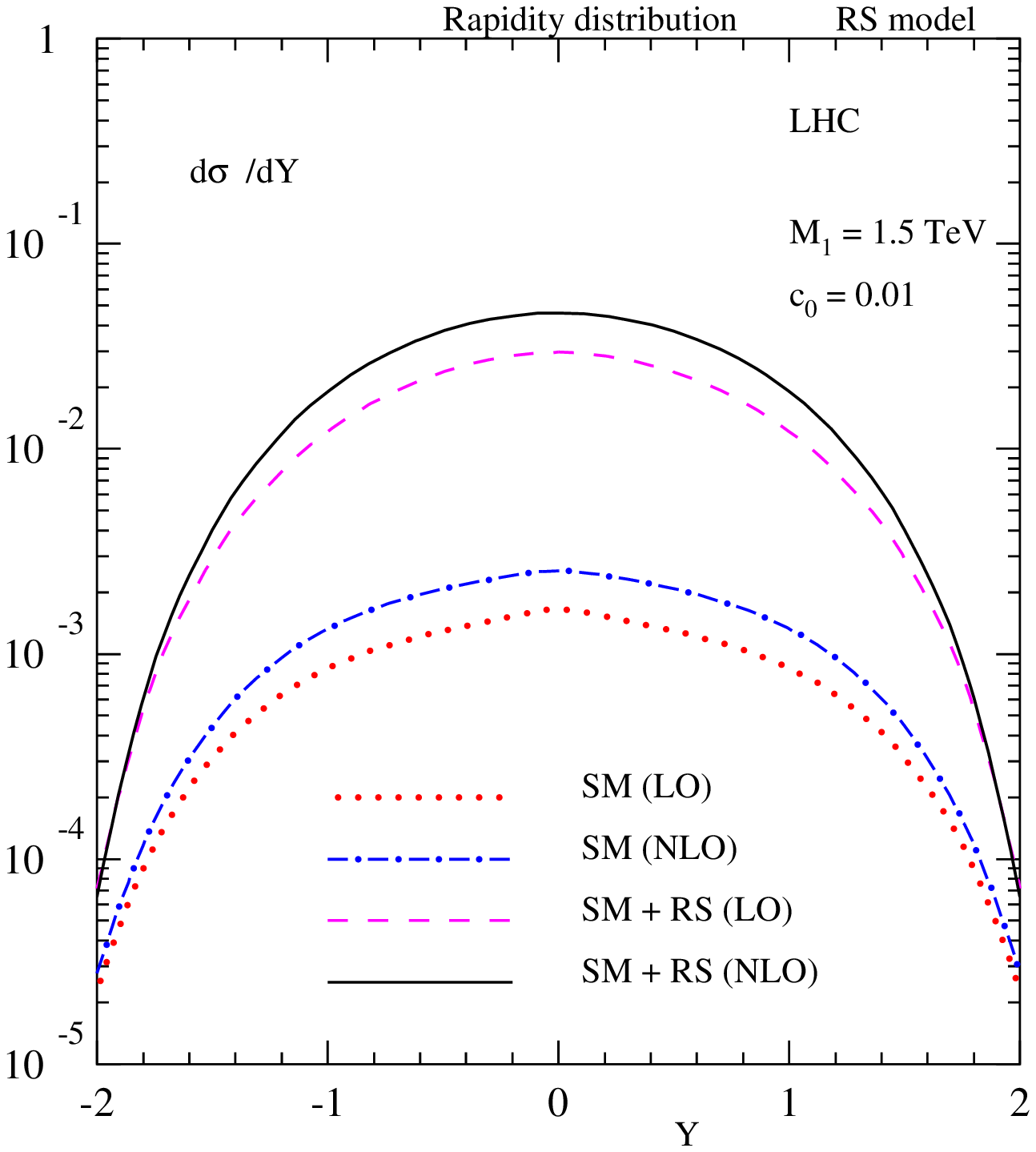,width=8cm,height=9cm,angle=0}}
\caption{Invariant mass $d\sigma/dQ$ (left) and rapidity $d\sigma/dY$ (right)
distributions of the di-photon production in the RS model with 
$M_1=1.5$ TeV and $c_0=0.01$ at the LHC.}
\label{rs-qf-yf}
\end{figure}

In this section we present the kinematic distributions of the photon pairs
in the RS model at the LHC to order $\alpha_s$.  Unlike the ADD model 
wherein the spectrum of the KK modes is uniform and almost continuous,
the spectrum in the RS model is quite non-uniform and contains heavy
resonances.  They can be probed via their resonance decays at large values of $Q$. 
In fig.(\ref{rs-qf-yf}), we present the invariant mass distribution 
of the di-photon in the RS model with the following choice of
parameters: (i) the mass of the first RS mode is $M_1=1.5$ TeV and 
(ii) the effective coupling between the RS modes and the SM fields is
$c_0=0.01$.  This choice is consistent
with the bounds obtained from the Tevatron \cite{Abazov:2007ra}.
The RS modes being heavy show up as resonances in the
invariant mass distribution which can be seen in the left panel of the figure.
In the right panel, we have plotted the rapidity distribution of the photon pairs
(see eqn.(\ref{eqn-Y})) after integrating the invariant mass of the photon pairs 
around the first resonance i.e. in the range $1100 < Q < 1600$ GeV.
We find that the signal has maximum contribution at the central rapidity (Y=0) 
and is differing from the SM background by an order of magnitude.  
The NLO QCD corrections enhance the signal and the background.
Even though, the resonance pattern in the high $Q$ region can point to new
physics, identification of the spin of the
resonance will be very important to discriminate between the various new physics
scenarios.  It is well known that spin information of these
resonances will be reflected in the angular distribution and we study them 
in the following.

\begin{figure}[htb]
\centerline{
\epsfig{file=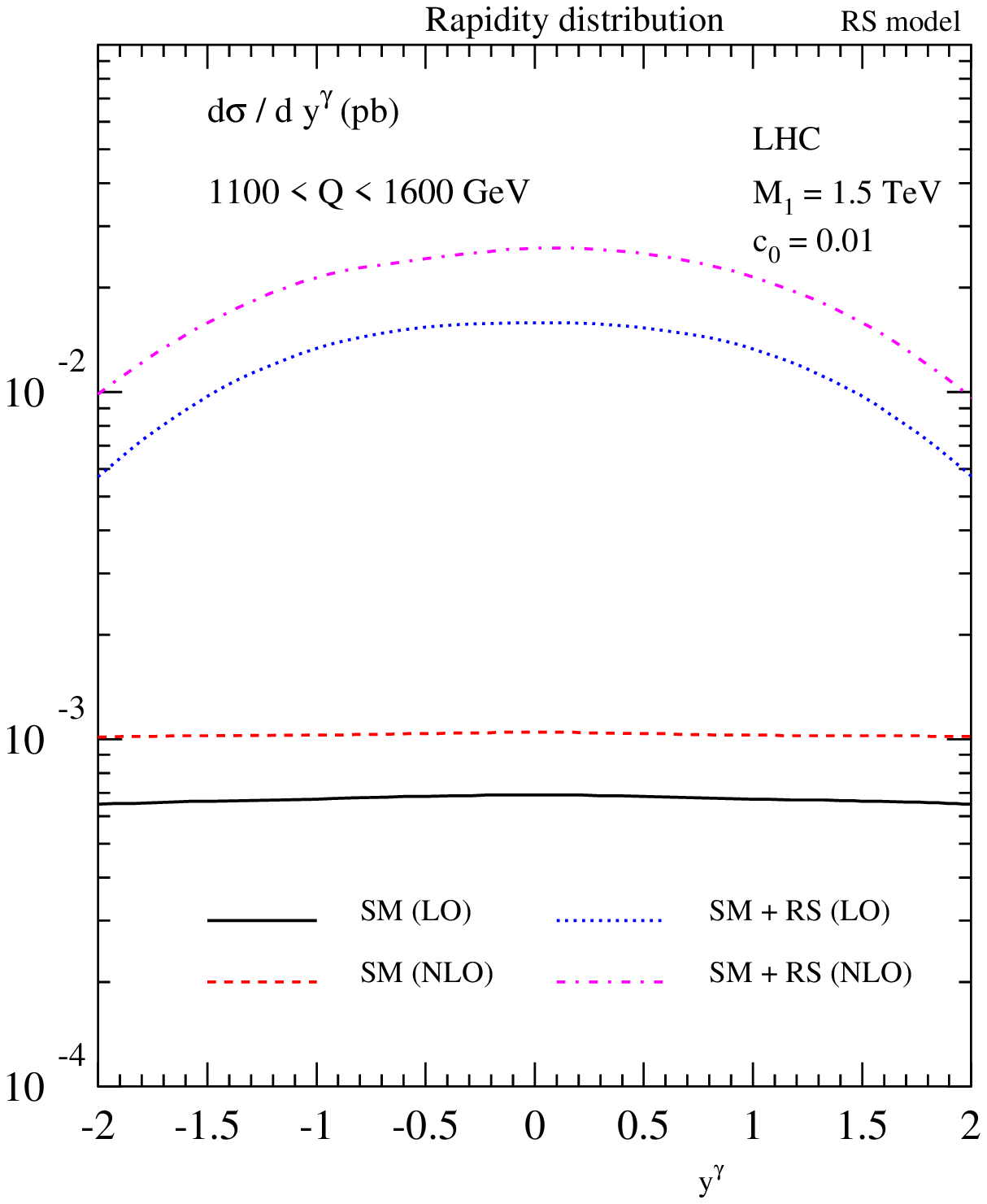,width=8cm,height=9cm,angle=0}
\epsfig{file=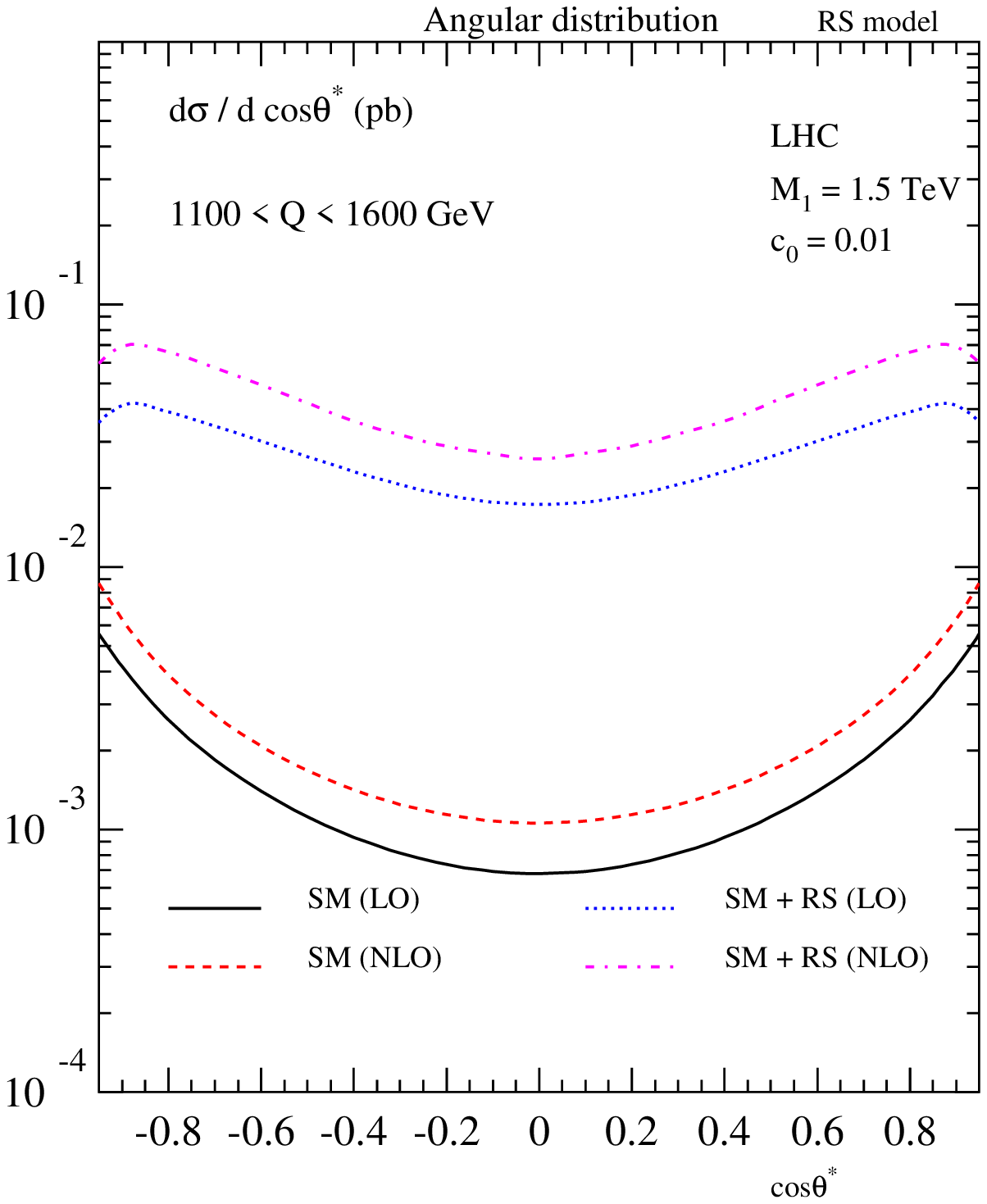,width=8cm,height=9cm,angle=0}}
\caption{Rapidity $d\sigma/dy^\gamma$ (left) and angular 
$d\sigma/d~\cos\theta^*$ (right) distributions of the photons in the RS 
model with $M_1=1.5$ TeV and $c_0=0.01$ at the LHC.}
\label{rs-y.ind-cf}
\end{figure}

\begin{figure}[htb]
\centerline{
\epsfig{file=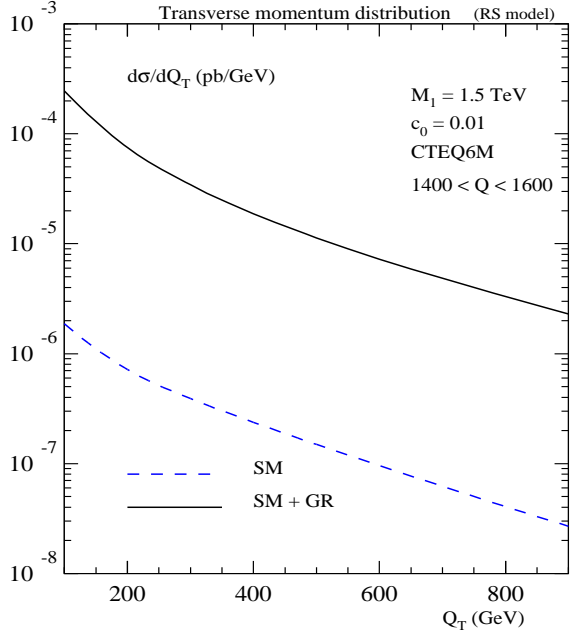,width=8cm,height=9cm,angle=0}}
\caption{Transverse momentum distribution of the di-photon
production in the RS model with $M_1=1.5$ TeV and $c_0=0.01$.
Here we have integrated over $Q$ around the first resonance 
region $1100 \le Q \le 1600$ GeV.}
\label{rs-qt}
\end{figure}

In fig.(\ref{rs-y.ind-cf}), rapidity $y^{\gamma}$ of the photons is plotted
$|y^{\gamma}| \le 2.0$ both in the SM and in SM+ADD to order $\alpha_s$. 
This distribution is obtained after integrating over the invariant mass of the photon pairs 
in the range $1100\le Q \le 1600$ GeV where RS resonance shows up.  
The SM cross sections show very little
variation with respect to $y^{\gamma}$ while the signal peaks at the central rapidity.
The cosine of the angle (see eqn.(\ref{eqn-cf})) between the final state 
photon and one of the incoming
hadrons in the c.o.m. frame of the final state photons 
is plotted  in the fig.(\ref{rs-y.ind-cf})
in the range $|\cos\theta^*| \le 0.95$. 
Again, we have restricted our $1100\le Q \le 1600$ GeV as in the case of $y^{\gamma}$
distribution.
The distribution coming from the SM has a minimum for the photons  
in the transverse direction and becomes large for the photons close
to the beam direction.
However, in the RS model, the signal shows an oscillating 
behavior and differs by more than an order of magnitude for the photons
in the transverse direction.  This is the feature unique to 
new physics scenarios where spin-2 objects decay to photon pairs.
We find our QCD corrections enhance the cross sections.
We present the $Q_T$ distribution which is non-zero only at order $\alpha_s$.
This is obtained after integrating 
$Q$ in the range $1100\le Q \le 1600$ GeV.  The numerical results are shown in
fig.(\ref{rs-qt}).  We find that the signal has
a large enhancement over the SM background for the entire range
of $Q_T$ considered  i.e. from 100 GeV to 900 GeV.
\subsection{Scale variations}
\begin{figure}[htb]
\centerline{
\epsfig{file=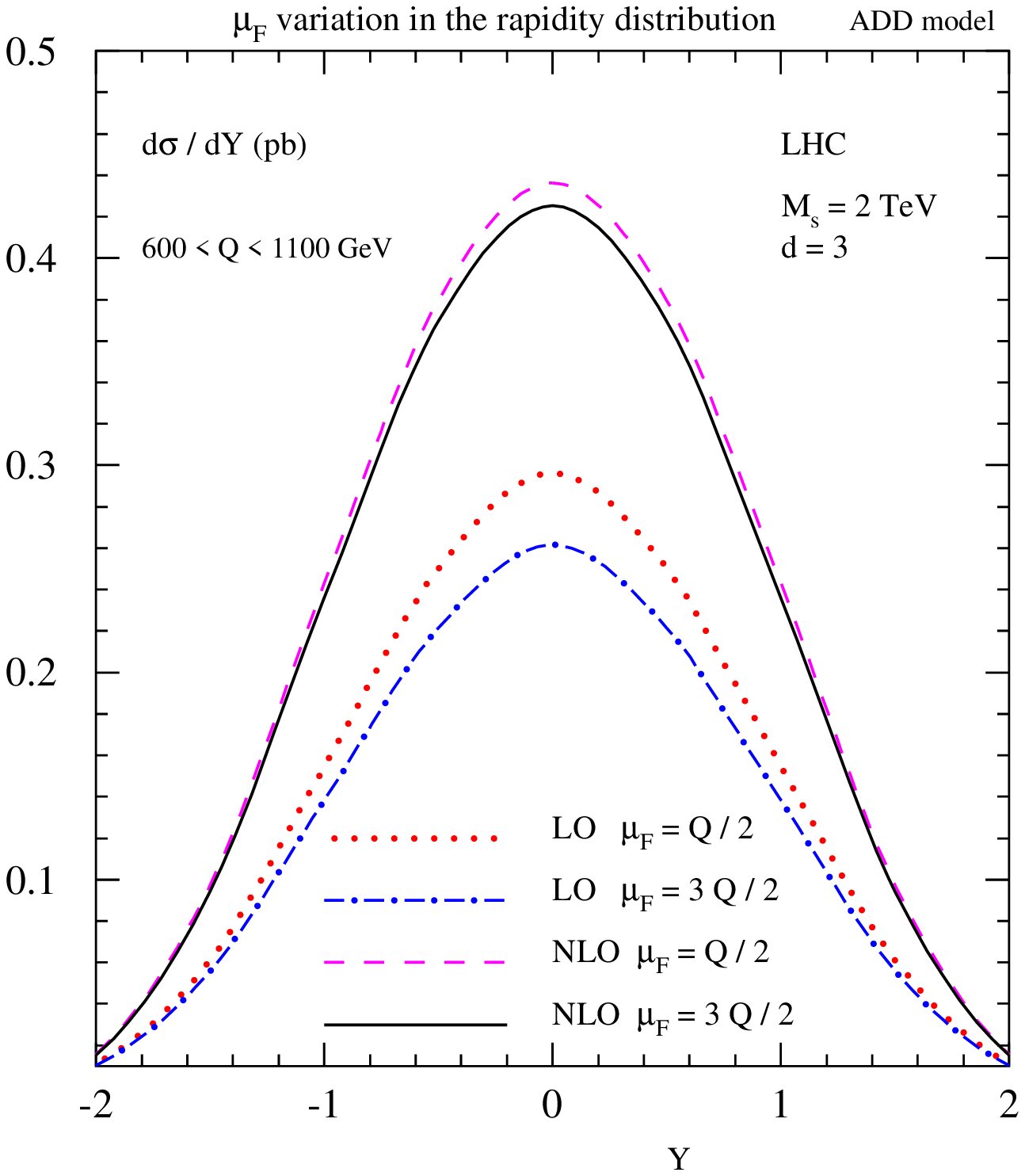,width=7.5cm,height=8.5cm,angle=0}
\epsfig{file=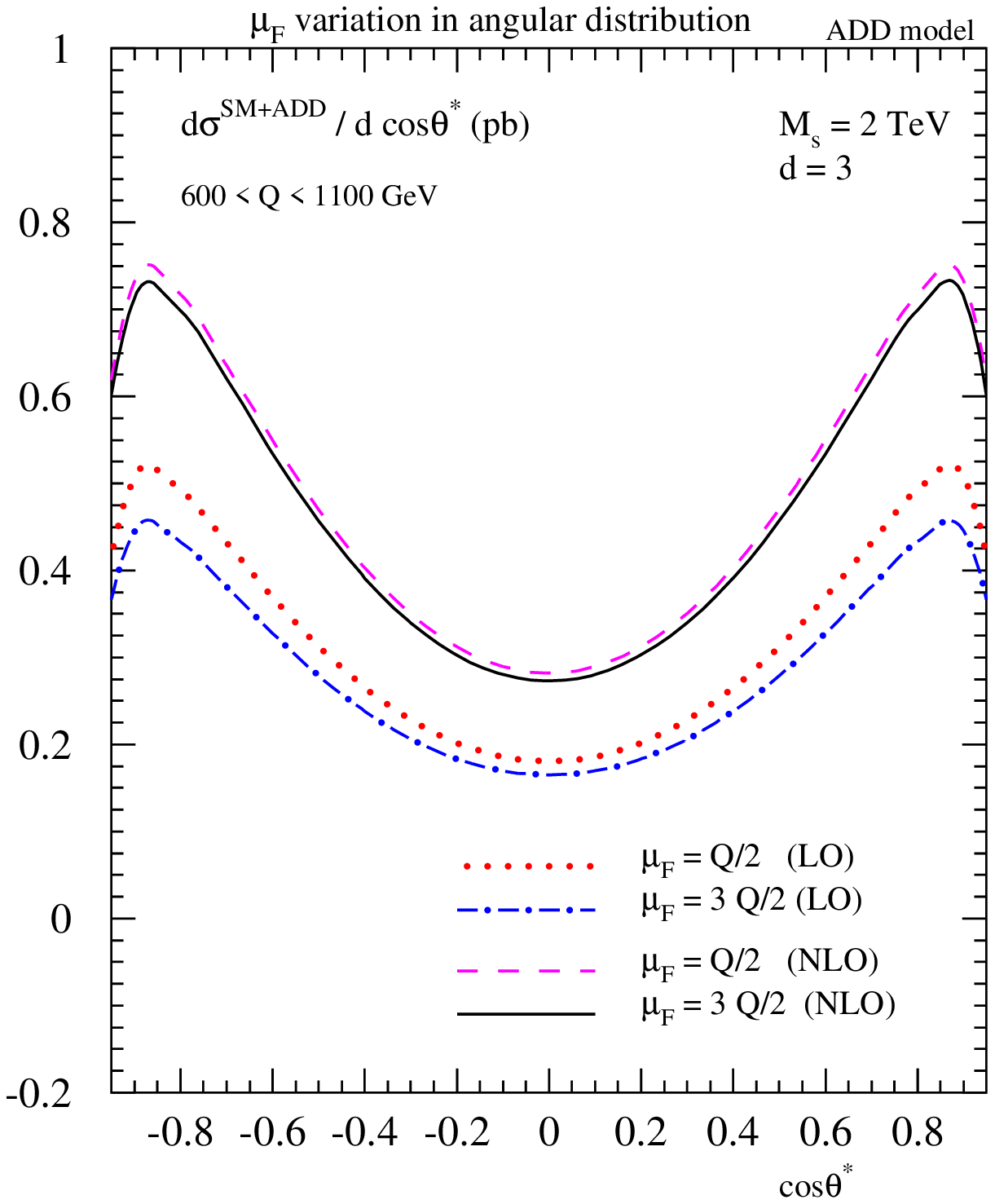,width=7.5cm,height=8.5cm,angle=0}}
\caption{Factorization scale dependency of the LO and NLO
cross sections in the ADD model with $M_S=2$ TeV and $d=3$
for a scale variation of $Q/2<\mu_F<3Q/2$.
For both the rapidity (left) and angular (right) distributions 
of the di-photon production, we have integrated over the invariant mass 
in the range $600~<Q~<1100$ GeV.}
\label{add-scale1}
\end{figure}

\begin{figure}[htb]
\centerline{
\epsfig{file=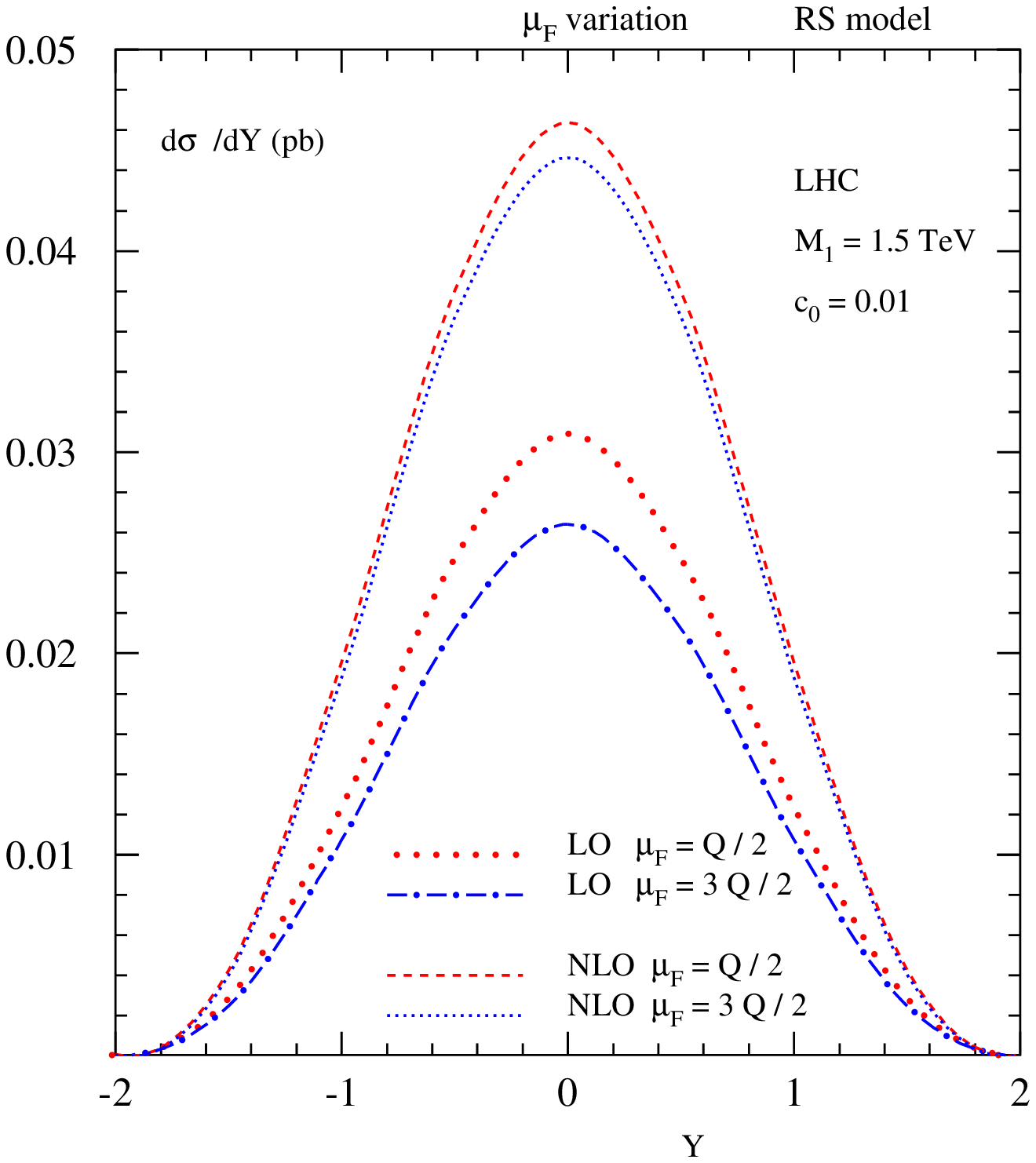,width=7.5cm,height=8.5cm,angle=0}}
\caption{The factorization scale dependency is shown in the 
rapidity distribution $d\sigma/dY$ of the di-photon for a scale 
variation of $Q/2 \le \mu_F \le 3Q/2$.  For this distribution
we have integrated over the invariant mass of the di-photon 
in the range $1100 \le Q \le 1600$ GeV.}
\label{rs-yf-muf}
\end{figure}

In this section, we discuss the impact of NLO QCD corrections to various
distributions.  The uncertainity in LO computation
of observables in the hadron colliders originates from two important sources,
namely, the missing higher order radiative corrections and the choice
of factorisation and renormalisation scales.  The former enters through
parton density sets and the latter through 
the renormalised parameters such as running coupling constant $\alpha_s$ 
of the theory.  The radiative corrections coming from QCD in our case enhance both SM 
as well as ADD and RS distributions.  Hence, the $K$-factor ($K=\sigma^{NLO}/\sigma^{LO}$), 
that quantifies these effects is always positive for the cases we studied in this paper.  
It is clear from the plots that
the $K$-factor is different for different distributions and also within a given distribution,
it varies with the kinematical variable, say $Q$ or $Y$ etc. 
More importantly, the numerical value of $K$ depends very much on the kinematical
cuts imposed on each distribution.  We find that the $K$ factors of the distributions
reported in the paper are not large and hence our NLO results are stable
under perturbation and reliable for further study.    
Observables are expected to be independent of renormalisation and factorisation scales, thanks to
renormalisation group invariance.  However, any truncated perturbative expansion
does depend on the choice of these scales.  This is expected to improve if higher
order corrections are included in the perturbative expansion.
Indeed, our NLO results of these distributions show significant 
improvement on the factorisation scale uncertainity entering through parton density sets at LO level.
In order that the perturbative expansion does not break down these scales should
be chosen close to the scale in the problem such as $Q$ or $Q_T$.  In the fig.(\ref{add-scale1})
we show the effect of variation of $\mu_F$ between $Q/2$ and $3 Q/2$.  We studied
this variation for $Y$ and $\cos\theta^*$ distributions in the ADD model.  A similar
analysis has been done for the RS model as well which is shown in fig.(\ref{rs-yf-muf})
Here, we have integrated the invariant mass around
the first resonance region $1100 \le Q \le 1600$ GeV.
\section{Conclusions}
In this paper, we have presented full next to leading order QCD corrections
to production of direct photon pairs at hadron colliders in the context
of extra-dimension scenarios namely ADD and RS models.
Both in ADD and RS models, photon pairs can be produced in a collision of partons through
virtual exchange of KK gravitons.  These give appreciable deviations to production rates
predicted in the SM due to large multiplicity of KK modes in the ADD and warp factor
in the RS model.  Only the spin-2 gravitons are included in our analysis and they
show distinct features in the angular distributions of photon pairs.  
Photon pairs at hadron colliders often provide a clean channel to probe
these physics beyond the SM.  In this study, only direct photons have been considered
and the fragmentation photons are removed by the method of smooth cone isolation.   
The isolation used also removes final state QED singularities.  
The leading order contributions to production rate resulting from quark and anti quark as well as
gluon initiated processes depend very much on the factorisation scale through
the parton distribution functions giving significant theory uncertainity.  In order
to bring down this uncertainity, we have systematically included all order $\alpha_s$
contributions to the process.  This includes all virtual and
real emission processes to order $\alpha_s$ both in SM as well as in ADD and RS models. 
To obtain various kinematic distributions of the final state photons, we have
used  phase space slicing method to deal with all the soft and collinear singularities
and the resulting finite pieces are integrated with the appropriate kinematical
cuts using a Monte Carlo program.  We have made several test on our NLO code
by showing the independence of the results on the slicing parameters and also 
comparing with the known NLO corrected SM results available in the literature.
Our numerical results including the NLO corrections show significant
enhancement over the LO predictions in all the distributions presented.
The enhancement varies with the distributions.  We have estimated them through
the K-factor which quantifies the reliability of the perturbative expansion.
We find the K-factor is moderate for all the distributions and hence the
results present here are stable under perturbation.  
We have also shown the impact of $\alpha_s$ corrected results on the scale
uncertainity.  We find that the factorisation scale dependence gets reduced
considerably when $\alpha_s$ contributions are included.

\noindent
{\bf Acknowledgments:}  
MCK would like to thank CSIR, New Delhi for financial support.  The work of VR and AT has 
been partially supported by funds made available to the Regional Centre 
for Accelerator-based Particle Physics (RECAPP) by the Department of 
Atomic Energy, Govt.  of India.  AT and VR would like to thank the  
cluster computing facility at Harish-Chandra Research Institute where 
part of computational work for this study was carried out.


\begin{figure}[h]
\SetScale{1.5}
\noindent
\begin{center}
\hspace{-5cm}
\begin{picture}(200,40)
\ArrowLine(0,40)(25,20)
\ArrowLine(25,20)(0,0) 
\DashLine(25,20)(50,20){4}
\Photon(50,20)(75,40){2}{3}
\Photon(50,20)(75,0){2}{3}
\Gluon(125,40)(150,20){2}{3}
\Gluon(150,20)(125,0){2}{3}
\DashLine(150,20)(175,20){4}
\Photon(175,20)(200,40){2}{3}
\Photon(175,20)(200,0){2}{3}
\end{picture}
\end{center}
\caption{Born contributions} 
\label{born}
\vspace{2cm}
\SetScale{1.5}
\noindent
\begin{center}
\hspace{-5cm}
\begin{picture}(200,40)
\ArrowLine(0,40)(20,40)
\ArrowLine(20,40)(20,0)
\ArrowLine(20,0)(0,0)
\Gluon(20,40)(38,20){2}{3}
\Gluon(38,20)(20,0){2}{3}
\DashLine(38,20)(55,20){4}
\Photon(55,20)(75,40){2}{3}
\Photon(55,20)(75,0){2}{3}
\ArrowLine(125,40)(150,20)
\ArrowLine(150,20)(125,0)
\Gluon(133,8)(133,33){2}{3}
\DashLine(150,20)(175,20){4}
\Photon(175,20)(200,40){2}{3}
\Photon(175,20)(200,0){2}{3}
\end{picture}
\end{center}
%
\vspace{1.0cm}
\begin{center}
\hspace{-5cm}
\begin{picture}(200,40)
\Gluon(0,40)(20,40){2}{3}
\ArrowLine(20,0)(20,40)
\Gluon(20,0)(0,0){2}{3}
\ArrowLine(20,40)(38,20)
\ArrowLine(38,20)(20,0)
\DashLine(38,20)(55,20){4}
\Photon(55,20)(75,40){2}{3}
\Photon(55,20)(75,0){2}{3}
\Gluon(125,40)(145,40){2}{3}
\ArrowLine(145,40)(145,0)
\Gluon(145,0)(125,0){2}{3}
\ArrowLine(163,20)(145,40)
\ArrowLine(145,0)(163,20)
\DashLine(163,20)(180,20){4}
\Photon(180,20)(200,40){2}{3}
\Photon(180,20)(200,0){2}{3}
\end{picture}
\end{center}
\vspace{1.0cm}
\begin{center}
\hspace{-5cm}
\begin{picture}(200,40)
\Gluon(0,40)(20,40){2}{3}
\DashArrowLine(20,0)(20,40){1}
\Gluon(20,0)(0,0){2}{3}
\DashArrowLine(20,40)(38,20){1}
\DashArrowLine(38,20)(20,0){1}
\DashLine(38,20)(55,20){4}
\Photon(55,20)(75,40){2}{3}
\Photon(55,20)(75,0){2}{3}
\Gluon(125,40)(145,40){2}{3}
\DashArrowLine(145,40)(145,0){1}
\Gluon(145,0)(125,0){2}{3}
\DashArrowLine(163,20)(145,40){1}
\DashArrowLine(145,0)(163,20){1}
\DashLine(163,20)(180,20){4}
\Photon(180,20)(200,40){2}{3}
\Photon(180,20)(200,0){2}{3}
\end{picture}
\end{center}
\vspace{1.0cm}
\begin{center}
\hspace{-5cm}
\begin{picture}(200,40)
\Gluon(0,40)(20,20){2}{3}
\Gluon(20,20)(0,0){2}{3}
\GlueArc(29,20)(9,0,360){2}{8}
\DashLine(38,20)(55,20){4}
\Photon(55,20)(75,40){2}{3}
\Photon(55,20)(75,0){2}{3}
\Gluon(125,40)(145,40){2}{3}
\Gluon(145,0)(145,40){2}{4}
\Gluon(145,0)(125,0){2}{3}
\Gluon(145,40)(163,20){2}{3}
\Gluon(163,20)(145,0){2}{3}
\DashLine(163,20)(180,20){4}
\Photon(200,40)(180,20){2}{3}
\Photon(180,20)(200,0){2}{3}
\end{picture}
\end{center}
\caption{Virtual contributions}
\label{virt}
\end{figure}


\begin{figure}[htb]
\centerline{\epsfig{file=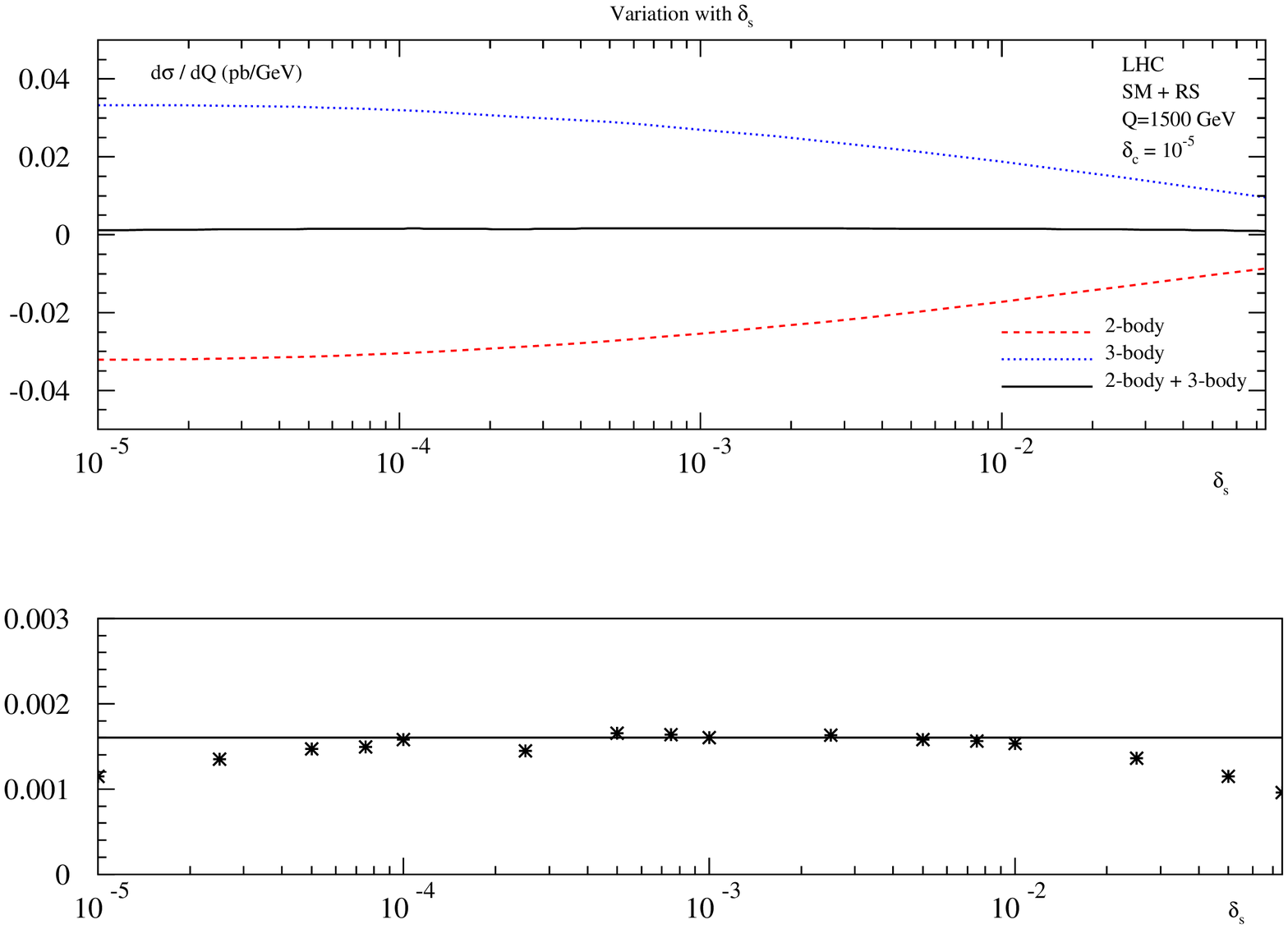,width=15cm,height=12cm,angle=0}}
\caption{Stability of the order $\alpha_s$ contribution to the SM+RS cross 
section against the variation of the slicing parameter $\delta_s$ (top),
 with $\delta_c = 10^{-5}$ fixed,  
in the invariant mass distribution of the di-photon with $M_1=1.5$ TeV and 
$c_0=0.01$.  Below is shown the variation of the sum of $2$-body and $3$-body
contributions over the range of $\delta_s$ considered and contrasted 
against the one at $\delta_s = 10^{-3}$.}
\label{rs-deltas}
\end{figure}

\begin{thebibliography}{99}
\bibitem{add}
  N.~Arkani-Hamed, S.~Dimopoulos and G.~R.~Dvali,
  Phys.\ Lett.\  B {\bf 429} (1998) 263;\\
I.\ Antoniadis, N.~Arkani-Hamed, S.~Dimopoulos and G.~R.~Dvali,
Phys.\ Lett.\  B {\bf 436} (1998) 257.
\bibitem{rs}
  L.~Randall and R.~Sundrum,
  Phys.\ Rev.\ Lett.\  {\bf 83} (1999) 3370;
W.D. Goldberger and M.B. Wise, Phys. Rev. Lett. {\bf 83} (1999) 4922.
\bibitem{expt}
C.\ D.\ Hoyle {\it et.~al}, {\it Phys. Rev.} {\bf D70} (2004) 042004.
\bibitem{hlz}
  T.~Han, J.~D.~Lykken and R.~J.~Zhang,
  Phys.\ Rev.\  D {\bf 59} (1999) 105006.
\bibitem{grw}
G.\ F.\  Giudice, R. Rattazzi and J. D. Wells, {\it Nucl. Phys.} {\bf B544}
(1999) 3.
\bibitem{peskin}
E.\ A.\ Mirabelli, M.\ Perelstein, M.\ E.\ Peskin
Phys.\ Rev.\ Lett.\ 82 (1999) 2236.
\bibitem{mrs}
Prakash Mathews, Sreerup Raychaudhuri, K. Sridhar
Phys.\ Lett.\ B450 (1999) 343; 
Prakash Mathews, Sreerup Raychaudhuri, K. Sridhar
Phys.\ Lett.\ B455 (1999) 115;
Prakash Mathews, Sreerup Raychaudhuri, K. Sridhar
JHEP 0007 (2000) 008.
\bibitem{gw} W.D. Goldberger and M.B. Wise, {\it Phys. Rev. Lett.}
{\bf 83} (1999) 4922; {\em ibid} {\it Phys.Lett.} {\bf B475} (2000) 275.
\bibitem{dhr} H. Davoudiasl, J.L. Hewett and T.G. Rizzo,
{\it Phys. Rev. Lett} {\bf 84} (2000) 2080; {\it ibid.} {\it Phys. Rev.}
{\bf D63} (2001) 075004.
\bibitem{Bern:2002jx}
  Z.~Bern, L.~J.~Dixon and C.~Schmidt,
  Phys.\ Rev.\  D {\bf 66} (2002) 074018
  [arXiv:hep-ph/0206194].
\bibitem{TwoPhotonBkgd1}
E.L.~Berger, E.~Braaten and R.D.~Field,
Nucl.\ Phys.\ B {\bf 239}, 52 (1984); 
P.~Aurenche, A.~Douiri, R.~Baier, M.~Fontannaz and D.~Schiff,
Z.\ Phys.\ C {\bf 29}, 459 (1985); 
B.~Bailey, J.F.~Owens and J.~Ohnemus,
Phys.\ Rev.\ D {\bf 46}, 2018 (1992); 
B.~Bailey and J.F.~Owens,
Phys.\ Rev.\ D {\bf 47}, 2735 (1993); 
B.~Bailey and D.~Graudenz,
Phys.\ Rev.\ D {\bf 49}, 1486 (1994)
[arXiv:hep-ph/9307368]; 
C.~Balazs, E.L.~Berger, S.~Mrenna and C.-P.~Yuan,
Phys.\ Rev.\ D {\bf 57}, 6934 (1998)
[arXiv:hep-ph/9712471]; 
C.~Balazs and C.-P.~Yuan,
Phys.\ Rev.\ D {\bf 59}, 114007 (1999)
[Erratum-ibid.\ D {\bf 63}, 059902 (1999)]
[arXiv:hep-ph/9810319];
T.~Binoth, J.P.~Guillet, E.~Pilon and M.~Werlen,
Phys.\ Rev.\ D {\bf 63}, 114016 (2001)
[arXiv:hep-ph/0012191]; 
T.~Binoth,
arXiv:hep-ph/0005194.
T.~Binoth, J.P.~Guillet, E.~Pilon and M.~Werlen,
Eur.\ Phys.\ J.\ C {\bf 16}, 311 (2000)
[arXiv:hep-ph/9911340].
\bibitem{BSMdiphoton}
  O.~J.~P.~Eboli, T.~Han, M.~B.~Magro and P.~G.~Mercadante,
  Phys.\ Rev.\  D {\bf 61} (2000) 094007
  [arXiv:hep-ph/9908358];
  K.~m.~Cheung and G.~L.~Landsberg,
  Phys.\ Rev.\  D {\bf 62} (2000) 076003
  [arXiv:hep-ph/9909218];
  M.~Luo, L.~Wang and G.~Zhu,
  Phys.\ Lett.\  B {\bf 672} (2009) 65
  [arXiv:0812.0866 [hep-ph]];
  M.~C.~Kumar, P.~Mathews, V.~Ravindran and A.~Tripathi,
  Phys.\ Lett.\  B {\bf 672} (2009) 45
  [arXiv:0811.1670 [hep-ph]].
\bibitem{drellyan}
  P.~Mathews, V.~Ravindran, K.~Sridhar and W.~L.~van Neerven,
  Nucl.\ Phys.\  B {\bf 713}, 333 (2005)
  [arXiv:hep-ph/0411018].
%
  P.~Mathews, V.~Ravindran and K.~Sridhar,
  JHEP {\bf 0510}, 031 (2005)
  [arXiv:hep-ph/0506158].
%
  P.~Mathews and V.~Ravindran,
  Nucl.\ Phys.\  B {\bf 753}, 1 (2006)
  [arXiv:hep-ph/0507250].
%
  M.~C.~Kumar, P.~Mathews and V.~Ravindran,
  Eur.\ Phys.\ J.\  C {\bf 49}, 599 (2007)
  [arXiv:hep-ph/0604135].
\bibitem{Abulencia:2006kk}
  A.~Abulencia {\it et al.}  [CDF Collaboration],
  Phys.\ Rev.\ Lett.\  {\bf 97} (2006) 171802
  [arXiv:hep-ex/0605101].
\bibitem{Abazov:2007ra}
  V.~M.~Abazov {\it et al.}  [D0 Collaboration],
  Phys.\ Rev.\ Lett.\  {\bf 100} (2008) 091802
  [arXiv:0710.3338 [hep-ex]].
\bibitem{ourpapers1}
  M.~C.~Kumar, P.~Mathews, V.~Ravindran and A.~Tripathi,
  Phys.\ Rev.\  D {\bf 77} (2008) 055013
  [arXiv:0709.2478 [hep-ph]];
  M.~C.~Kumar, P.~Mathews, V.~Ravindran and A.~Tripathi,
  arXiv:0804.4054 [hep-ph];
\bibitem{ourpapers2}
  M.~C.~Kumar, P.~Mathews, V.~Ravindran and A.~Tripathi,
  Phys.\ Lett.\  B {\bf 672} (2009) 45
  [arXiv:0811.1670 [hep-ph]].
\bibitem{FORM} 
FORM by J.A.M.~Vermaseren, version 3.0 available from http://www.nikhef.nl/form.
\bibitem{twocutoff}
L.J.~Bergmann, Next-to-leading-log QCD calculation of symmetric dihadron production,
Ph.D. thesis, Florida State University, 1989; L. Bergmann and J.F.~Owens, Report No. FSU-HEP-890601
(unpublished).
\bibitem{Bailey:1992br}
  B.~Bailey, J.~F.~Owens and J.~Ohnemus,
  Phys.\ Rev.\  D {\bf 46} (1992) 2018.
\bibitem{twocutoff-rev}
  B.~W.~Harris and J.~F.~Owens,
  Phys.\ Rev.\  D {\bf 65} (2002) 094032
  [arXiv:hep-ph/0102128].
\bibitem{gg}
  R.~K.~Ellis, I.~Hinchliffe, M.~Soldate and J.~J.~van der Bij,
  Nucl.\ Phys.\  B {\bf 297} (1988) 221;
L.~Ametller, E.~Gava, N.~Paver and D.~Treleani,
Phys.\ Rev.\ D {\bf 32}, 1699 (1985); \\
%
D.A.~Dicus and S.S.D.~Willenbrock,
Phys.\ Rev.\ D {\bf 37}, 1801 (1988).
\bibitem{Frixione:1998jh}
  S.~Frixione,
  Phys.\ Lett.\  B {\bf 429} (1998) 369.
\bibitem{atlas}
ATLAS Collaboration, {\rm ATLAS detector and physics performance.
Technical design report.}  Vol. 2 (1999), CERN-LHCC-99-15.

\bibitem{cms}
CMS Collaboration, "CMS: The electromagnetic calorimeter, technical design report," 
report CERN/LHCC 97-33, CMS-TDR-4.
\bibitem{Pumplin:2002vw}
  J.~Pumplin {\it et al.},
  JHEP {\bf 0207} (2002) 012.
\end{thebibliography}
\end{document}